\documentclass[11pt,a4paper]{article}
\pdfoutput=1
\usepackage{amssymb}
\usepackage{amsmath}
\usepackage{graphicx}
\usepackage{booktabs}
\usepackage{color}
\usepackage{textcomp}
\usepackage{multirow}
\usepackage{bm}
\usepackage{caption}
\usepackage{subcaption}
\usepackage{tikz}
\usepackage{longtable}
\usepackage{colortbl}
\usepackage[utf8]{inputenc}

\usepackage{jheppub}

\begin{document}


\title{Excited and exotic charmonium, $D_s$ and $D$ meson spectra for two light quark masses from lattice QCD}

\author[a]{Gavin~K.~C.~Cheung,} \emailAdd{gkcc2@damtp.cam.ac.uk}
\author[b]{Cian~O'Hara,} \emailAdd{oharaci@tcd.ie}
\author[a]{Graham~Moir,} \emailAdd{graham.moir@damtp.cam.ac.uk}
\author[b]{Michael~Peardon,} \emailAdd{mjp@maths.tcd.ie}
\author[b]{Sin\'{e}ad~M.~Ryan,} \emailAdd{ryan@maths.tcd.ie}
\author[a]{Christopher~E.~Thomas,} \emailAdd{c.e.thomas@damtp.cam.ac.uk}
\author[b]{David~Tims} \emailAdd{timsd@tcd.ie}

\affiliation[a]{DAMTP, University of Cambridge, Centre for Mathematical Sciences, Wilberforce Road, Cambridge CB3 0WA, UK}
\affiliation[b]{School of Mathematics, Trinity College, Dublin 2, Ireland}

\affiliation{\vspace{0.1cm} {\rmfamily \normalsize (for the Hadron Spectrum Collaboration)}}

\abstract{We present highly-excited charmonium, $D_s$ and $D$ meson spectra from dynamical lattice QCD calculations with light quarks corresponding to $M_{\pi} \sim 240$ MeV and compare these to previous results with $M_{\pi} \sim 400$ MeV. Utilising the distillation framework, large bases of carefully constructed interpolating operators and a variational procedure, we extract and reliably identify the continuum spin of an extensive set of excited mesons.  These include states with exotic quantum numbers which, along with a number with non-exotic quantum numbers, we identify as having excited gluonic degrees of freedom and interpret as hybrid mesons.  Comparing the spectra at the two different $M_\pi$, we find only a mild light-quark mass dependence and no change in the overall pattern of states.}

\preprint{DAMTP-2016-63}


\maketitle

\section{Introduction}\label{sec:introduction}

The experimental status of the charm sector of Quantum Chromodynamics (QCD) has changed dramatically over the last decade \cite{PDG2015}.  The discovery of a plethora of unexpected charmonium-like states, commonly known as ``$X, Y, Z$'s'', has highlighted the need for a more complete theoretical understanding of the spectrum.  Many different interpretations have been put forward: some are suggested to be hybrid mesons (a quark-antiquark pair with excited gluonic degrees of freedom) and others two quarks and two antiquarks in a tightly-bound configuration (tetra-quark), a molecular-like combination of two mesons, or a charmonium-like core surrounded by light degrees of freedom (hadro-quarkonium).  There are similar puzzles in the open-charm sector ($D$ and $D_s$ mesons) where the measured masses and widths of the low-lying $D^{*}_{s0}(2317)^{\pm}$ and $D_{s1}(2460)^{\pm}$ states are significantly smaller and narrower than expected from quark models.  For some recent reviews see Refs.~\cite{Brambilla:2010cs,Brambilla:2014jmp,Olsen:2015zcy,Swanson:2015wgq,Prencipe:2015kva}.

In principle these states can be understood within Quantum Chromodynamics (QCD) using lattice QCD, a non-perturbative, \textit{ab initio} formulation of the theory.  Spurred on by the experimental situation, there have been many lattice QCD calculations of hidden and open-charm mesons. The majority have focused on lowest-lying states below threshold, achieving unprecedented precision with the various systematic effects under control (some recent examples can be found in Refs.~\cite{Namekawa:2011wt, McNeile:2012qf, Dowdall:2012ab, Donald:2012ga, Galloway:2014tta}). On the other hand, there have been a number of investigations of excited charmonia and open-charm mesons~\cite{Dudek:2007,Bali:2011rd, Bali:2011dc, Mohler:2011ke, Bali:2015lka, Kalinowski:2015bwa, Cichy:2016bci}, all of which have some systematic uncertainties not fully accounted for and extract a more limited set of states than we consider here.

In a previous lattice QCD study, the Hadron Spectrum Collaboration used large bases of interpolating operators with various structures to robustly extract many excited and high-spin states and, crucially, to identify their continuum quantum numbers.  Highlights included the presence of states with exotic quantum numbers (i.e.\ those forbidden with solely a quark-antiquark pair) and the identification of ``supermultiplets'' of hybrid mesons.  However, these calculations were performed with unphysically-heavy light quarks corresponding to $M_{\pi} \sim 400$ MeV.  The results provided useful benchmarks for other approaches such as nonrelativistic effective field theories, for example see Ref.~\cite{Berwein:2015vca}.

The current work extends these earlier investigations by performing similar calculations with light-quark masses significantly closer to their physical values, corresponding to $M_{\pi} \sim 240$ MeV.  The spectra at the two light quark masses are compared, focusing on the overall qualitative picture and, in particular, whether changes in the pattern of states with exotic quantum numbers or other hybrid mesons are observed.  This allows us to explore the light-quark mass dependence of excited heavy quarkonia which has been suggested to be significant~\cite{Guo:2012tg}.

In this study the unstable nature of states above threshold is not considered -- a point discussed in~\cite{Dudek:2010,Liu:2012,Moir:2013ub} -- and so the spectra should only be considered a guide to the pattern of resonances.  In the charm sector, we have already addressed this limitation for a variety of states appearing as bound-states and resonances in coupled-channel $D\pi$, $D\eta$ and $D_{s}\bar{K}$ scattering~\cite{Moir:2016srx} for $M_{\pi} \sim 400$ MeV and investigations of various other channels involving charm quarks are underway.  This paper lays the foundation for extending those studies to $M_{\pi} \sim 240$ MeV, where the additional light-quark mass, closer to the physical value, will enable us to study the evolution with light-quark mass of hidden and open-charm bound-states and resonances.

A number of other investigations of near-threshold bound states, scattering and resonances in the charm sector have appeared over the last few years~\cite{Ozaki:2012ce, Mohler:2012na, Liu:2012zya, Prelovsek:2013cra, Prelovsek:2013xba, Mohler:2013rwa, Chen:2014afa, Lang:2014yfa, Prelovsek:2014swa, Padmanath:2015era, Lang:2015sba, Chen:2015jwa, Chen:2016lkl,Ikeda:2016zwx}.  There have also been studies addressing the existence of four-quark configurations (mostly considering static heavy quarks)~\cite{Bicudo:2012qt, Brown:2012tm, Ikeda:2013vwa, Bicudo:2015vta, Bicudo:2015kna, Francis:2016hui, Alberti:2016dru, Peters:2016isf,Bicudo:2016jwl}. However, these are mainly exploratory and more comprehensive calculations as described in Ref.~\cite{Moir:2016srx} are called for.

The remainder of the manuscript is organised as follows.  In Section~\ref{sec:calculation_details} we describe the lattice ensembles used in this study, provide some details on the tuning of the anisotropy and charm-quark mass, and give a brief overview of the analysis of two-point correlation functions.  In Section~\ref{sec:spectra} we present and interpret the charmonium, $D_s$ and $D$ meson spectra from the calculations with $M_{\pi} \sim 240$ MeV.  In Section~\ref{sec:comparison} we compare these spectra to those from earlier computations with $M_{\pi} \sim 400$ MeV and we present a summary in Section~\ref{sec:conclusions}.

\section{Calculation Details}\label{sec:calculation_details}

\begin{table}[tb]
\begin{center}
\begin{tabular}{|c|c|c|c|c|}
\hline
 Lattice Volume & $M_{\pi}$ (MeV) & $N_{\rm cfgs}$ & $N_{\rm tsrcs}$ for $c\bar{c}$, $c\bar{s}$, $c\bar{l}$ & $N_{\rm vecs}$ \\
\hline
 $24^{3}\times 128$ & 391 & 553 & 32, 16, 16 & 162 \\
\hline
 $32^3\times 256$ & 236 &  484 & 1, 1, 2 & 384  \\
\hline
\end{tabular}
\caption{The lattice gauge field ensembles used.  The volume is given as
$(L/a_s)^3 \times (T/a_t)$ where $L$ and $T$ are respectively the spatial and temporal extents of the lattice.  The number of gauge field configurations used, $N_{\rm cfgs}$, and the number
of perambulator time-sources used per configuration, $N_{\rm tsrcs}$, are shown along with the number of eigenvectors used in the distillation framework~\cite{Peardon:2009}, $N_{\rm vecs}$.}
\label{tab:lattice_details}
\end{center}
\end{table}

In this study we use an anisotropic lattice formulation where the temporal lattice spacing, $a_t$, is smaller than the spatial lattice spacing, $a_s \approx 0.12$~fm, with an anisotropy $\xi \equiv a_s / a_t \approx 3.5$.  The gauge sector is described by a tree-level Symanzik-improved anisotropic action, while the fermionic sector uses a tadpole-improved anisotropic Sheikholeslami-Wohlert (clover) action with stout-smeared gauge fields~\cite{Morningstar:2003} and $N_{f}=2+1$ flavours of dynamical quarks.  For both ensembles the heavier dynamical quark is tuned to approximate the physical strange quark, but the ensembles differ in light quark mass giving the two different pion masses.  Table~\ref{tab:lattice_details} summarises these lattice ensembles -- full details are given in Refs.~\cite{Edwards:2008,Lin:2009}.

\begin{figure}[tb]
\begin{minipage}{.5\linewidth}
\includegraphics[width=0.95\textwidth]{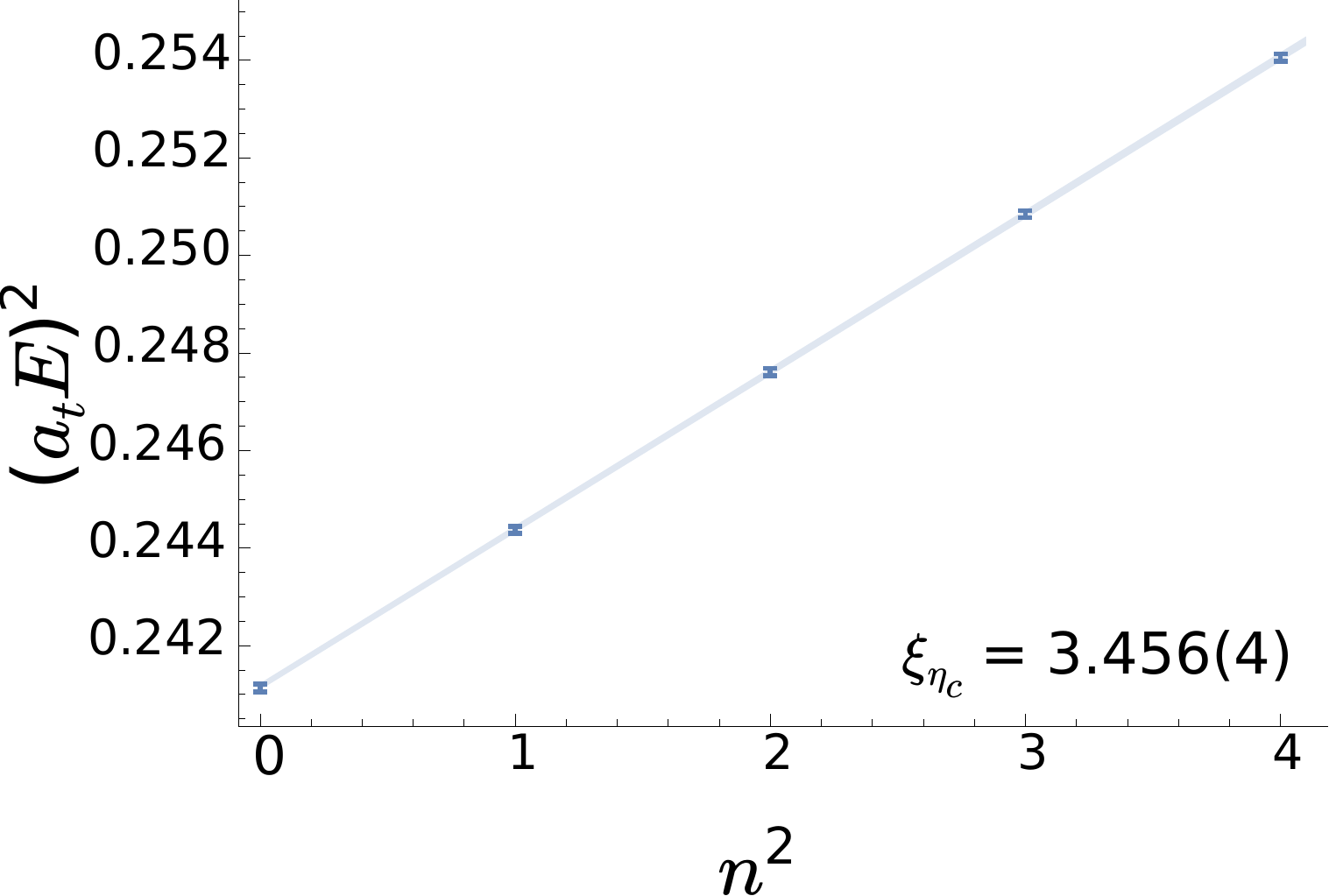}
\end{minipage}
\begin{minipage}{.5\linewidth}
\includegraphics[width=0.95\textwidth]{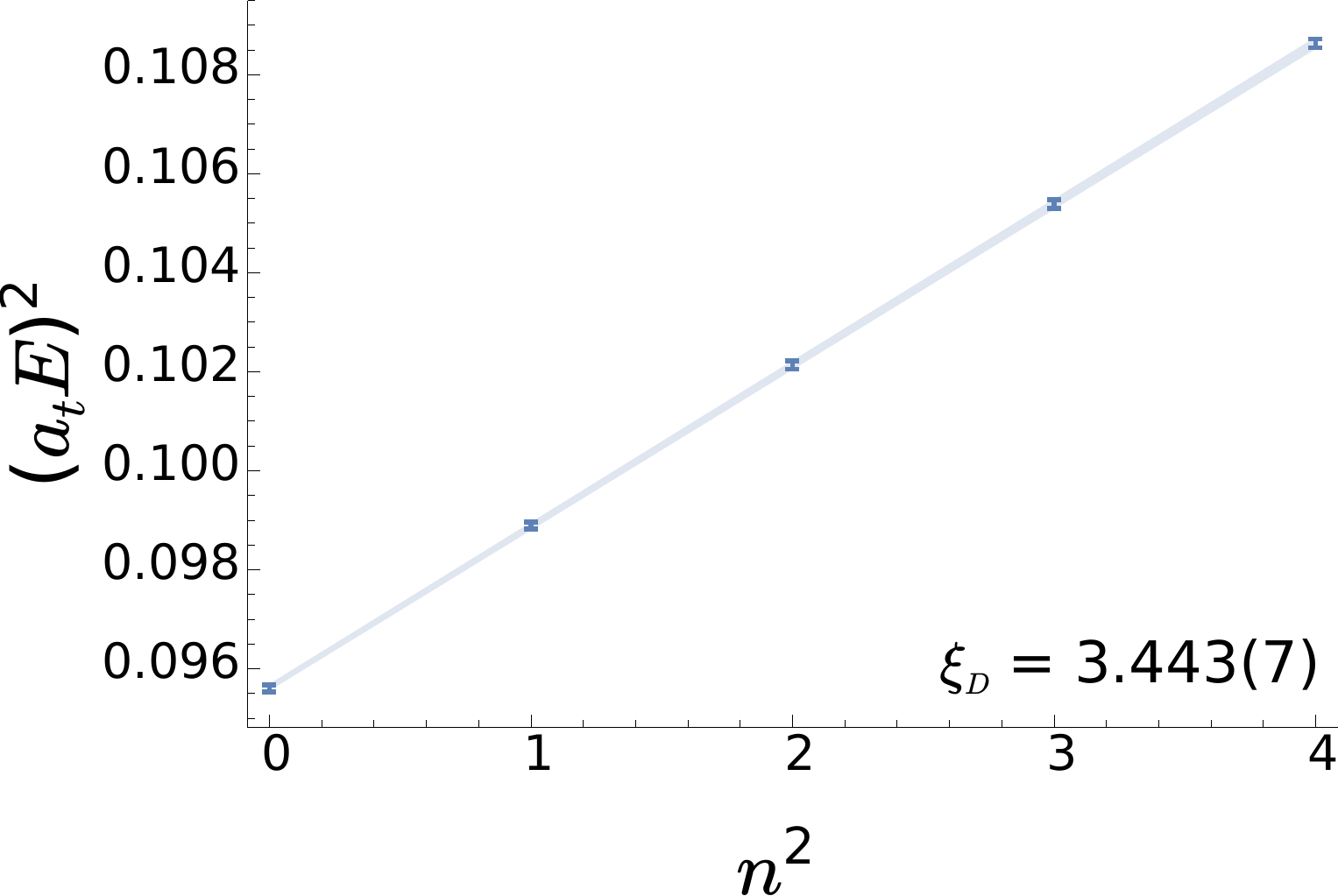}
\end{minipage}
\caption{Points show the dependence of the $\eta_c$ (left panel) and $D$ (right panel) energy on momentum; error bars show the one sigma statistical uncertainty on either side of the mean.  Lines are fits to the relativistic dispersion relation, Eq.~\ref{equ:dispersion}, giving $\xi_{\eta_c} = 3.456(4)$ ($\chi^{2}/N_\mathrm{d.o.f} = 1.08$) and $\xi_D = 3.443(7)$ ($\chi^{2}/N_\mathrm{d.o.f} = 0.38)$.}
\label{fig:dispersion}
\end{figure}

We use the same relativistic action for the charm quark as for the light and strange quarks (with tadpole-improved tree-level clover coefficients). The charm-quark mass and anisotropy parameters are tuned to reproduce the physical $\eta_c$ mass and a relativistic dispersion relation -- this process was described for the $M_{\pi} \sim 400$ MeV ensemble in Ref.\ \cite{Liu:2012}. Throughout this work we do not correct experimental data for electromagnetic effects. For the $M_{\pi} \sim 240$ MeV ensemble, the momentum dependence of the $\eta_c$ energy after tuning is shown in Figure~\ref{fig:dispersion}.  The momentum is quantised by the periodic boundary conditions on the cubic spatial volume, $\vec{p} = \frac{2\pi}{L} \vec{n}$, where $\vec{n}=(n_x,n_y,n_z)$ and $n_i \in \{0, 1, 2, \dots, L/a_s - 1 \}$.  A reasonable fit to the dispersion relation,
\begin{equation}
\label{equ:dispersion}
(a_t E)^2 = (a_t M)^2 + \left( \frac{2\pi}{\xi L/a_s} \right)^2 n^2 \, ,
\end{equation}
is obtained giving $\xi_{\eta_c} = 3.456(4)$, in agreement with the anisotropy measured from the pion dispersion relation on this ensemble, $\xi_{\pi} = 3.453(6)$~\cite{Wilson:2015dqa}.
The fit gives $M_{\eta_c} = 2945(17)$ MeV compared to the experimental value $2983.6(6)$~MeV~\cite{PDG2015}, and so we estimate that the systematic uncertainty from tuning the charm-quark mass is of order 1\%.
Figure~\ref{fig:dispersion} also shows the momentum dependence of the $D$ meson energy; a fit to Eq.~\ref{equ:dispersion} gives $\xi_{D} = 3.443(7)$, in reasonable agreement with $\xi_{\pi}$ and $\xi_{\eta_c}$.

To give results in physical units, we set the scale via $a_t^{-1} = M_\Omega^{\mathrm{phys}} / (a_t M_\Omega)$ using the $\Omega$ baryon mass measured on this ensemble, $a_t M_\Omega = 0.2789(16)$~\cite{Wilson:2015dqa}, leading to $a_t^{-1} = 5997$ MeV. When quoting masses we reduce the already small systematic uncertainty from tuning the charm-quark mass by subtracting $M_{\eta_c}$ ($\tfrac{1}{2} M_{\eta_c}$) from the mass of charmonia (open-charm mesons), rendering it negligible compared to other systematic uncertainties.

The aim of this work is to study how the spectra change as we vary the light-quark mass and only statistical uncertainties are given in the spectra we present in the following sections. While a full error budget is beyond the scope of this work, the uncertainties arising from working at a finite lattice spacing and in a finite volume were discussed in Ref.~\cite{Liu:2012}, where they were estimated to be small and have no overall qualitative effect on the spectrum. The uncertainty arising from the ambiguity in how to set the scale can be estimated by choosing a different reference observable. For example, setting the scale on the $M_\pi \sim 240$ MeV ensemble using the $h_c$ -- $\eta_c$ mass splitting gives $a_t^{-1} = 5960$ MeV, $0.6\%$ lower than from using $M_\Omega$. On the other hand, using the $\eta_c(2S)$ -- $\eta_c(1S)$ mass splitting gives $a_t^{-1} = 5787$ MeV, $4\%$ lower than when using the $\Omega$ baryon mass. 

Another source of systematic uncertainty comes from ignoring the unstable nature of states above threshold (see Refs.~\cite{Dudek:2010,Liu:2012,Moir:2013ub}). Although this is difficult to estimate, for a narrow resonance a conservative approach is to consider the uncertainty to be of the order of the width~\cite{Dudek:2010}.

\subsection{Calculation of spectra}

We follow the methodology presented in Refs.~\cite{Dudek:2010,Liu:2012,Moir:2013ub} to compute the spectra.  In brief, meson masses and other spectral information are obtained from the analysis of the time dependence of two-point Euclidean correlation functions,
\begin{equation}
C_{ij}(t) = \langle 0 | \mathcal{O}_i(t) \mathcal{O}^\dagger_j(0) | 0 \rangle~,
\end{equation}
where $\mathcal{O}^\dagger(0)$ [$\mathcal{O}(t)$] is the creation operator [annihilation operator] and $t$ is the time separation.  When computing charmonium correlators, disconnected Wick diagrams, where the charm quark and antiquark annihilate, are not included -- these are OZI suppressed and so are expected to only give a small contribution in charmonium.  There are no such disconnected contributions to the open-charm meson correlators considered here.

The hypercubic lattice has a reduced symmetry compared to an infinite volume continuum so states at rest are labelled by the irreducible representations (\emph{irreps}), $\Lambda$, of the octahedral group, $O_h$, rather than spin~\cite{Johnson:1982yq}.  
A method to ameliorate this issue and determine the continuum spin, $J$, of extracted states is given in Refs.~\cite{Dudek:2010,Liu:2012,Moir:2013ub} which also contain demonstrations of its efficacy.  Parity, $P$, and any relevant flavour quantum numbers, e.g.\ charge-conjugation, $C$, are still good quantum numbers in our lattice formulation.

In each quantum-number channel, the distillation technique~\cite{Peardon:2009} is used to compute correlation functions involving a large basis of derivative-based fermion-bilinear interpolating operators~\cite{Dudek:2010}.\footnote{To investigate more completely the resonant nature of states above threshold we would need to supplement the basis with operators of additional structures, e.g.~multi-meson operators, as in Ref.~\cite{Moir:2016srx}.}  The resulting matrices of correlation functions, $C_{ij}(t)$, are analysed using a variational procedure~\cite{Michael:1985,Luscher:1990,Blossier:2009kd} as described in Ref.~\cite{Dudek:2010}. This amounts to solving a generalised eigenvalue problem,
$C_{ij}(t) v^{\mathfrak{n}}_j = \lambda^{\mathfrak{n}}(t,t_0) C_{ij}(t_0) v^{\mathfrak{n}}_j $, where $t_{0}$ is a carefully chosen reference time-slice.  For sufficiently large times, the eigenvalues, $\lambda^{\mathfrak{n}}(t,t_0)$, known as principal correlators, are proportional to $e^{-M_{\mathfrak{n}}(t-t_{0})}$ where $M_{\mathfrak{n}}$ is the energy of the $\mathfrak{n}^{th}$ state.  
Energies are extracted from a fit to the form, $(1 - A_{\mathfrak{n}}) e^{-M_{\mathfrak{n}}(t-t_0)} + A_\mathfrak{n} e^{-M'_{\mathfrak{n}} (t-t_0)}$, 
where the fit parameters are $M_{\mathfrak{n}}$, $A_{\mathfrak{n}}$ and $M'_{\mathfrak{n}}$.  The second exponential proves useful in stabilising the fit because it `mops up' excited state contamination.
The eigenvectors, $v^{\mathfrak{n}}_{j}$, are related to the operator-state overlaps (or matrix elements), $Z_i^{(\mathfrak{n})} \equiv \langle \mathfrak{n} | \mathcal{O}_i^\dagger | 0 \rangle$, and contain information on the structure of a state -- they are used in our method for determining the continuum spin.

\begin{figure}[tb]
\begin{center}
\includegraphics[width=0.9\textwidth]{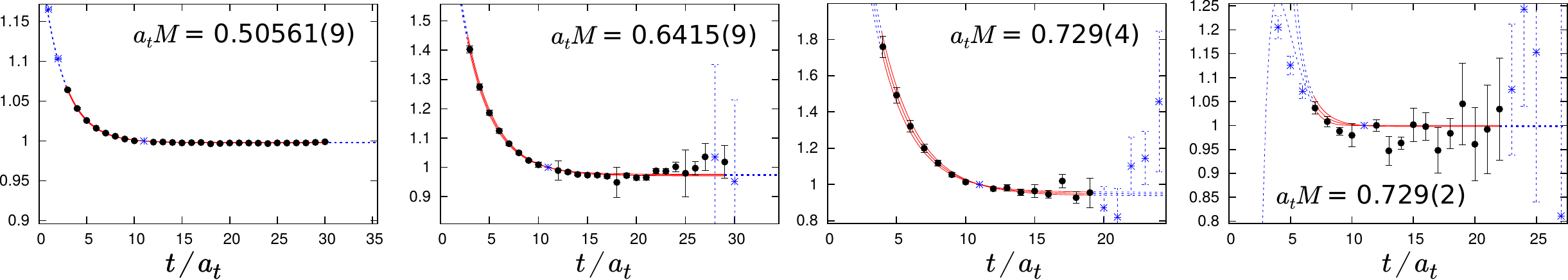}
\includegraphics[width=0.9\textwidth]{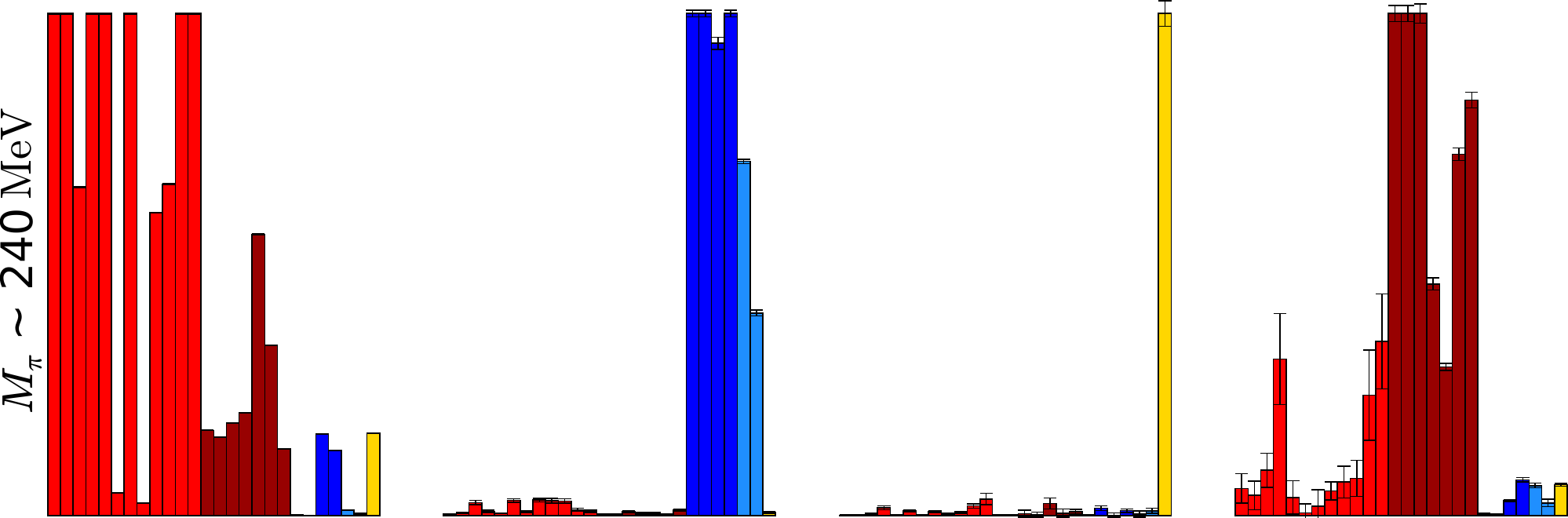}
\includegraphics[width=0.9\textwidth]{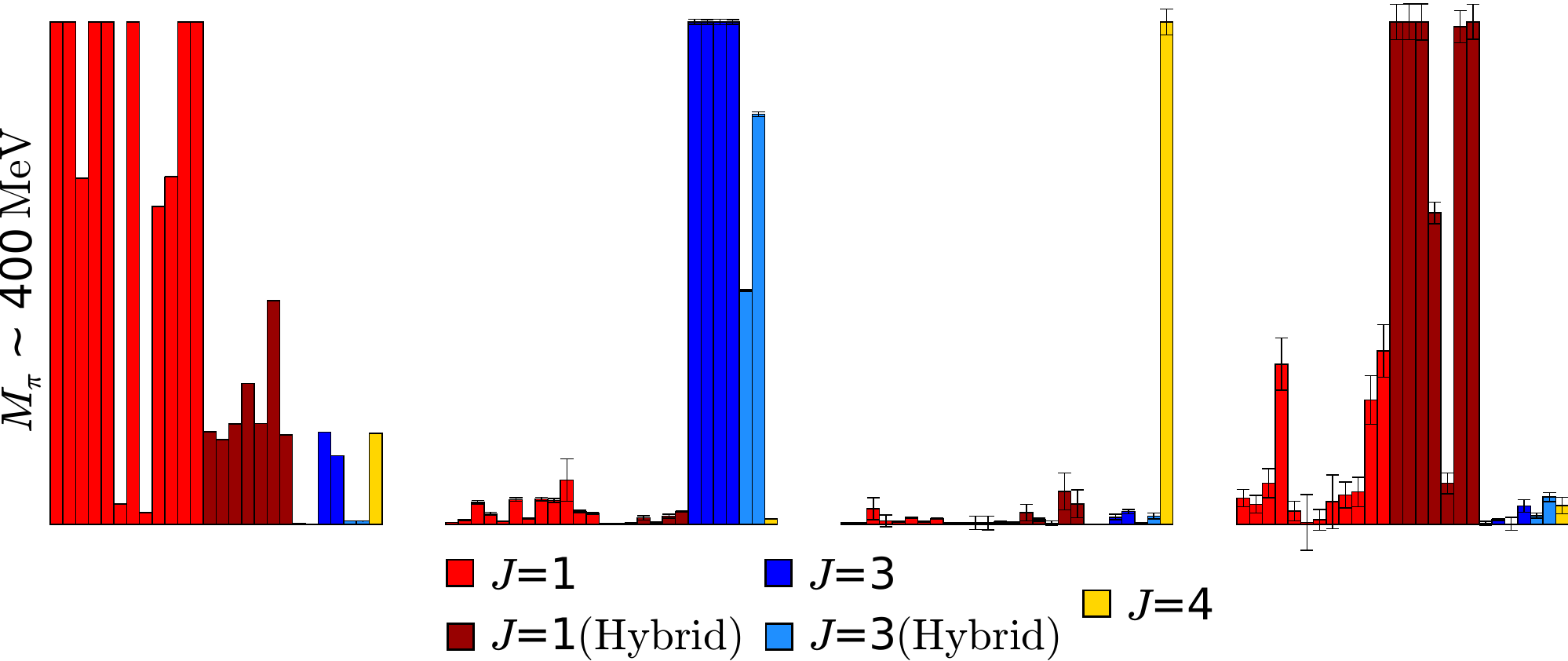}
\end{center}
\caption{Top row: principal correlators for a selection of low-lying charmonium states in the $T_1^{--}$ irrep on the $M_\pi \sim 240$ MeV ensemble. The data (points) and fits (curves) for $t_0 = 11$ are plotted as $\lambda^{(\mathfrak{n})} e^{M_\mathfrak{n}(t-t_0)}$ showing the central values and one sigma statistical uncertainties.  In each case the fit is reasonable with $\chi^2/N_\mathrm{d.o.f} \sim 1$.  Red parts of the curves show the time regions used in the fits; blue points were not included in the fits.  
Middle row: the operator-state overlaps, $Z$, for the state above, normalised so that the largest value for an operator across all states is equal to unity.  Colour coding is described in the text and the error bars indicate the one sigma statistical uncertainty.
Bottom row: overlaps for the corresponding state on the $M_\pi \sim 400$ MeV ensemble.}
\label{fig:prin_corrs}
\end{figure}

Figure~\ref{fig:prin_corrs} shows a selection of principal correlators from charmonium correlation functions in the $\Lambda^{PC} = T_{1}^{--}$ irrep on the $M_\pi \sim 240$ MeV ensemble.  The leading time dependence, $e^{-M_{\mathfrak{n}}(t-t_0)}$, has been divided out yielding a plateau when a single exponential dominates.
Beneath each principal correlator we show the overlap, $Z$, of each operator onto that state and below that, for comparison, the overlaps for the corresponding state on the $M_\pi \sim 400$ MeV ensemble.  The operators were constructed to have definite $J^{PC}$ in the continuum: red bars correspond to $J=1$, blue to $J=3$ and yellow to $J=4$.  It is clear that each state is dominated by operators from a given $J$, demonstrating that the spin-identification methodology~\cite{Dudek:2010} can be used -- this pattern is repeated for each of the spectra we determine. The darker shade of red and lighter shade of blue represent operators that are proportional to the spatial part of the field strength tensor, $F_{ij}$. We identify a state as hybrid, i.e. a meson with excited gluonic degrees of freedom~\cite{Dudek:2010}, when overlaps from these operators onto a given state are large compared to their overlaps onto other states\footnote{In Figure~\ref{fig:prin_corrs} the apparently considerable overlap of the $J=3$ state with hybrid operators is an artefact of the normalisation; in absolute terms these overlaps are small and we do not identify that state as a hybrid.}.

\section{Charmonium and Open-Charm Spectra}
\label{sec:spectra}

In this section we present the spectra, labelled by $J^{P(C)}$, computed on the $M_{\pi} \sim 240$ MeV ensemble.  
Results for charmonium are described first followed by those for $D_s$ and $D$ mesons.

\subsection{Charmonium}
\label{sec:spectra:charmonium}

\begin{figure}[tb]
\begin{center}
     \includegraphics[width=0.99\textwidth]{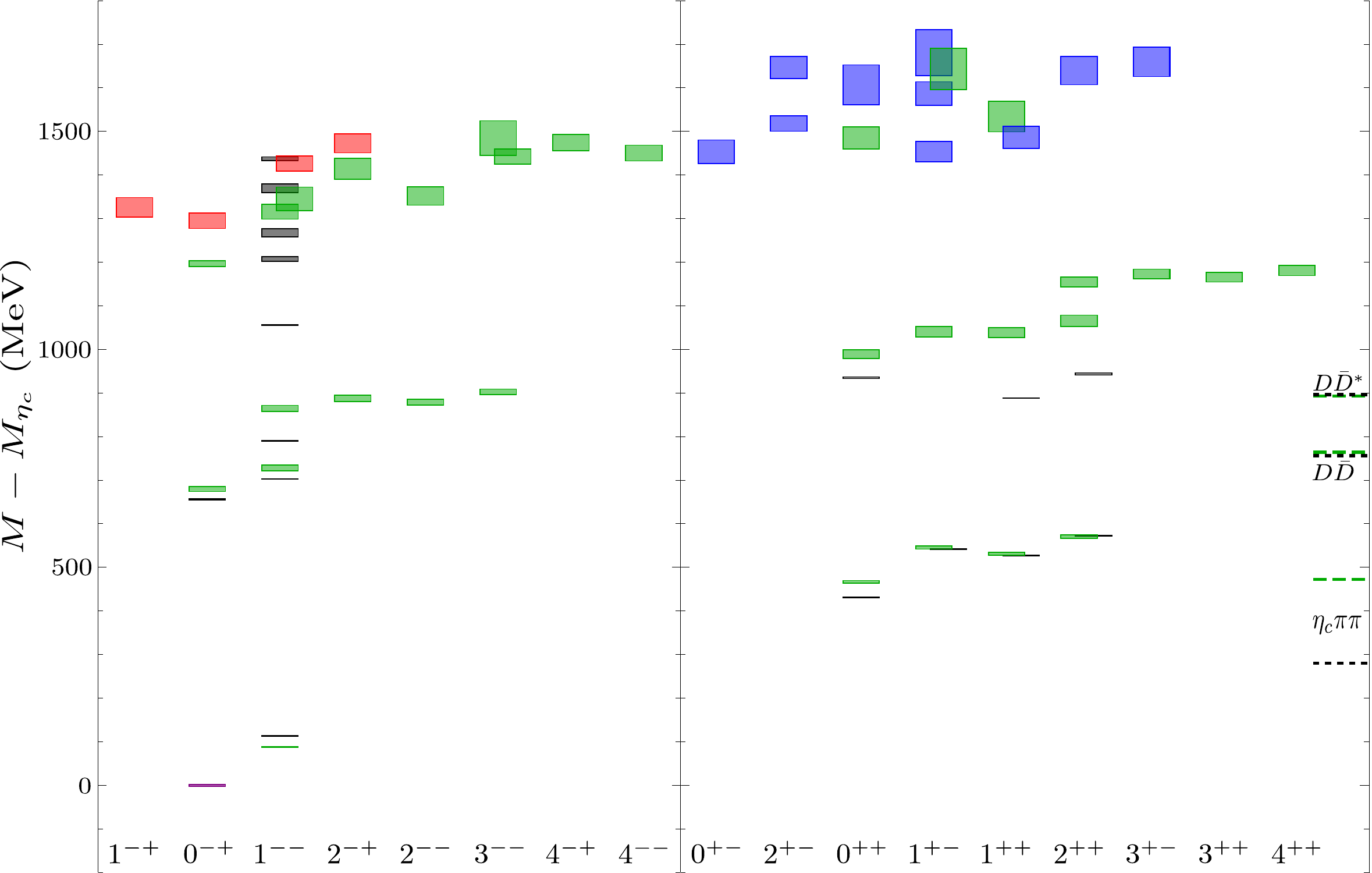}
\end{center}
    \caption{Charmonium spectrum up to around $4.5$ GeV labelled by $J^{PC}$; the left (right) panel shows the negative (positive) parity states.  Green, red and blue boxes are the masses computed on our $M_\pi\sim 240$ MeV ensemble while black boxes are experimental values from the PDG summary tables~\cite{PDG2015}. As discussed in the text, we show the calculated (experimental) masses with the calculated (experimental) $\eta_c$ mass subtracted.  The vertical size of the boxes represents the one-sigma statistical (or experimental) uncertainty on either side of the mean.  Red and blue boxes correspond to states identified as hybrid mesons grouped into, respectively, the lightest and first-excited supermultiplet, as described in the text.  Dashed lines show the location of some of the lower thresholds for strong decay using computed (coarse green dashing) and experimental (fine grey dashing) masses.}   
  \label{fig:charmonium_spectrum}
\end{figure}

The charmonium spectrum computed on the $M_\pi\sim 240$ MeV ensemble is shown in Figure~\ref{fig:charmonium_spectrum} and the results are tabulated in 
Appendix~\ref{app:tables}.
For flavour singlets such as charmonium, charge-conjugation, $C$, and parity, $P$, are both good quantum numbers and so states are labelled by $J^{PC}$.  As discussed above, masses are presented after subtracting the $\eta_c$ mass to reduce the systematic uncertainty arising from tuning the charm quark mass. 
Dashed lines indicate the location of some thresholds for strong decay: $\eta_c \pi \pi$ (the lowest threshold if the charm quark and antiquark do not annihilate), $D\bar{D}$ and $D\bar{D}^*$.  Since the resonant nature of states above threshold is not investigated in this work, a conservative approach is to only consider the mass values accurate up to the order of the hadronic width~\cite{Dudek:2010}.

As found in Ref.~\cite{Liu:2012}, many of the states with non-exotic $J^{PC}$ follow the $n^{2S+1}L_{J}$ pattern predicted by quark potential models, where $J$ is the total spin of the meson with relative orbital angular momentum $L$, quark-antiquark spin $S$ and radial quantum number $n$.  
We find all states up to $J=4$ expected by such models.

Figure~\ref{fig:charmonium_spectrum} also shows the states (coloured red and blue) that do not fit the $n^{2S+1}L_{J}$ pattern.  Four of these have exotic $J^{PC}$ quantum numbers, $0^{+-}, 1^{-+}, 2^{+-}$, and we find that they, as well as the excess states with non-exotic quantum numbers, have relatively large overlaps onto operators that are proportional to the spatial components of the field strength tensor, $F_{ij}$ (i.e. operators that have a non-trivial gluonic structure), something not seen for the other states in the spectrum. Furthermore, on removing operators proportional to $F_{ij}$ from the variational basis we generally observe a reduction in the quality of the signal for these states. We therefore follow Refs.~\cite{Dudek:2010,Liu:2012} and interpret these excess states as hybrid mesons.

As discussed in detail in Ref.~\cite{Liu:2012}, the hybrid states can be grouped into supermultiplets.  We find that the set $\left[(0^{-+}, 1^{-+}, 2^{-+}), 1^{--}\right]$, highlighted in red in Figure~\ref{fig:charmonium_spectrum}, forms the lightest charmonium hybrid supermultiplet, while the
states highlighted in blue, $(0^{++}, 1^{++}, 2^{++})$, $(0^{+-}, 1^{+-}, 1^{+-}, 1^{+-}, 2^{+-}, 2^{+-}, 3^{+-})$, form the first excited hybrid supermultiplet. These patterns are consistent with a quark-antiquark pair coupled to a $1^{+-}$ gluonic excitation; the lightest hybrid supermutiplet has the quark-antiquark pair in $S$-wave and the first excited hybrid supermultiplet has it in $P$-wave.
The lightest hybrids appear $\sim 1.2$ - $1.3$ GeV above the lightest $S$-wave meson multiplet.
This pattern of hybrids and their energy scale are consistent with what was found in the light meson and baryon sectors~\cite{Dudek:2010,Dudek:2011b,Dudek:2012,Edwards:2012fx,Dudek:2013yja}, studies of charmed baryons~\cite{Padmanath:2013zfa,Padmanath:2015jea} and in our previous work on charmonia and open-charm mesons~\cite{Liu:2012,Moir:2013ub}.

As noted in Section~\ref{sec:calculation_details}, these calculations are performed at a single spatial lattice spacing. On the $400$ MeV ensemble we estimated a scale of $40$ MeV for the discretisation uncertainty arising from $\mathcal{O}(a_{s})$ corrections to charmonia~\cite{Liu:2012}. Since the $240$ MeV ensemble has the same spatial lattice spacing, we expect the 40 MeV scale to also be a reasonable estimate for the discretisation uncertainty here.

\subsection{$D_s$ and $D$ mesons}
\label{sec:spectra:opencharm}

\begin{figure}[tb]
\begin{center}
\includegraphics[width=0.99\textwidth]{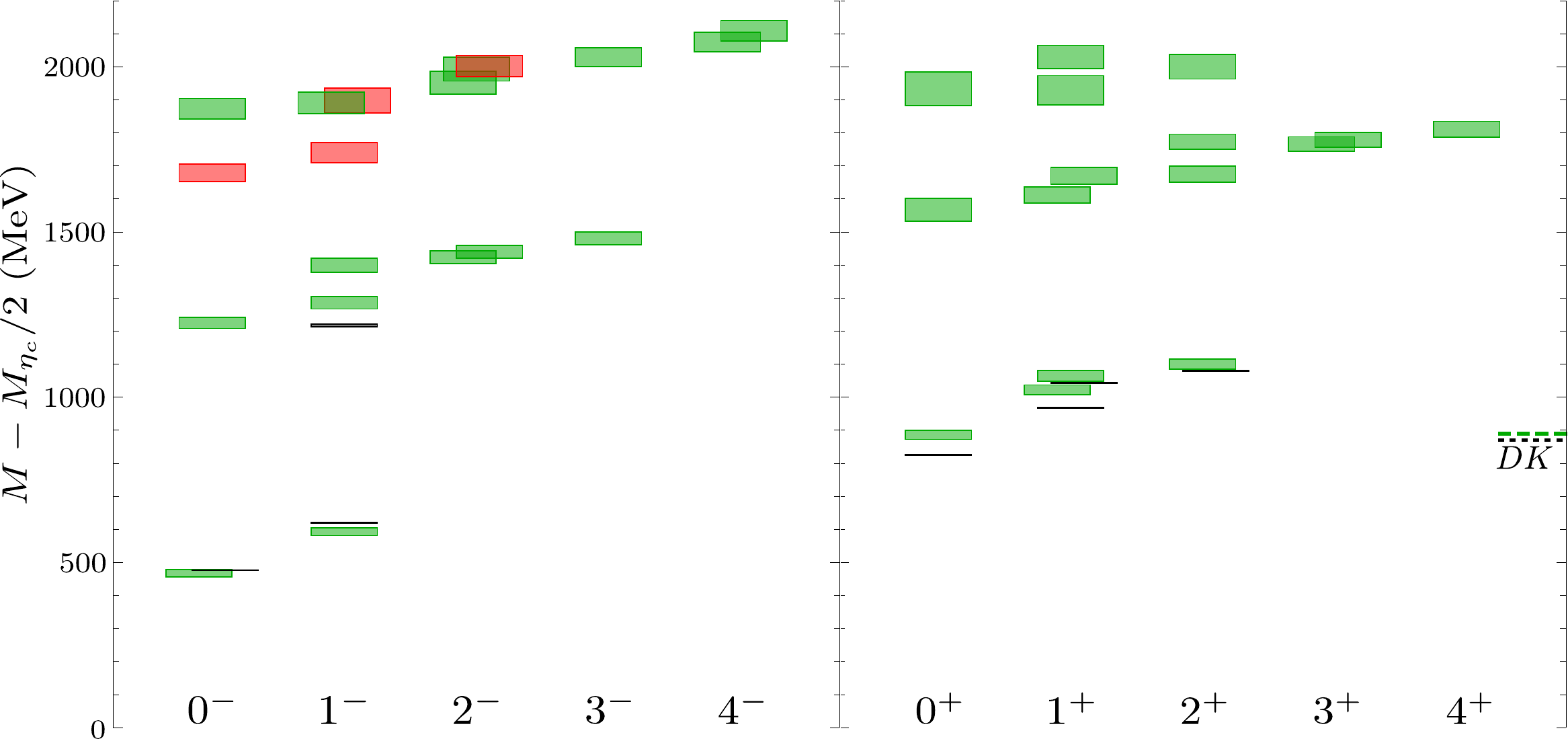}
\caption{$D_s$ meson spectrum labelled by $J^P$; the left (right) panel shows the negative (positive) parity states.  Green and red boxes are the masses computed on the $M_\pi\sim 240$ MeV ensemble while black boxes are experimental masses of the neutral $D$ mesons from the PDG summary tables~\cite{PDG2015}.  As discussed in the text, the calculated (experimental) masses are shown with with half the calculated (experimental) $\eta_c$ mass subtracted.  The vertical size of the boxes indicates the one-sigma statistical (or experimental) uncertainty on either side of the mean.  Red boxes show states identified as constituting the lightest hybrid supermultiplet, as described in the text.  
Dashed lines indicate the $DK$ threshold using computed (coarse green dashing) and experimental (fine grey dashing) masses.}
\label{fig:Ds_spectrum}
\end{center}
\end{figure}

For flavoured mesons, such as $D_{s}$ and $D$, charge conjugation is no longer a good quantum number and states are labelled only by $J^{P}$. Figures~\ref{fig:Ds_spectrum} and \ref{fig:D_spectrum} show the $D_s$ and $D$ meson spectra respectively; these results are tabulated in Appendix~\ref{app:tables}.
Masses are presented with half the mass of the $\eta_c$ subtracted in order to reduce the systematic uncertainty arising from tuning the charm quark mass.
Dashed lines indicate some of the lower strong-decay thresholds ($DK$ for the $D_s$ spectrum and $D\pi$ and $D^{*}\pi$ for the $D$ meson spectrum).

As for charmonium, the $D_s$ and $D$ spectra can be interpreted in terms of a $n^{2S+1}L_{J}$ pattern and we identify complete $S,P,D$ and $F$-wave multiplets.
Within the negative parity sector of both spectra, there are four states, highlighted in red, that do not appear to fit this pattern.  Due to their relatively large overlap with operators featuring a non-trivial gluonic structure, these are identified as the members of the lightest hybrid meson supermultiplet in each flavour sector.  The pattern is again consistent with a $1^{+-}$ gluonic excitation coupled to an $S$-wave quark-antiquark pair and they appear at an energy $\sim 1.2$ - $1.3$ GeV above the lightest conventional multiplet.  However, unlike in charmonium, the first excited hybrid supermultiplet is not 
robustly determined for open-charm mesons and is not shown here. 

\begin{figure}[tb]
\begin{center}
\includegraphics[width=0.99\textwidth]{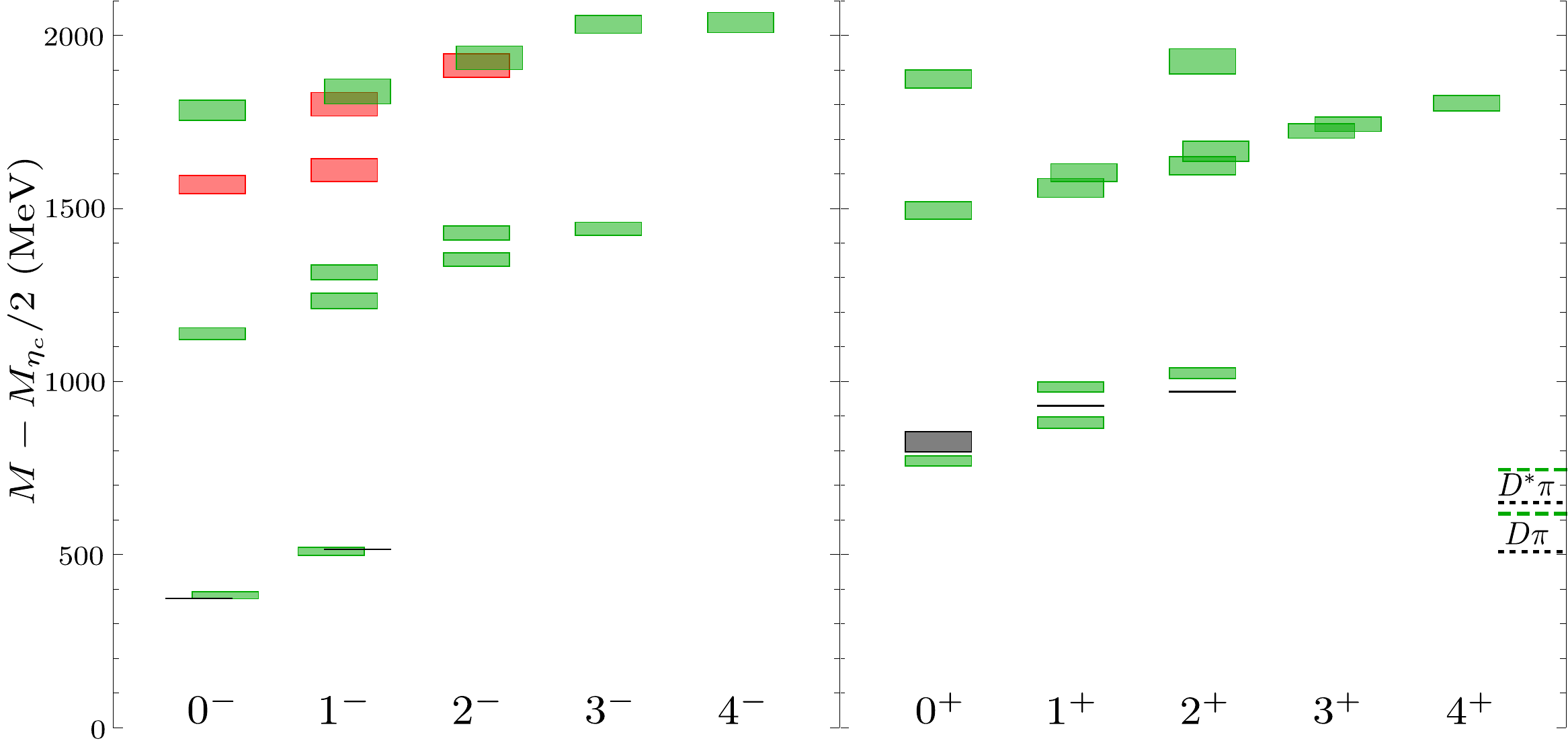}
\caption{As Figure~\ref{fig:Ds_spectrum} but for the $D$ meson spectrum.
Dashed lines show the $D\pi$ and $D^\ast\pi$ thresholds using computed (coarse green dashing) and experimental (fine grey dashing) masses.}
\label{fig:D_spectrum}
\end{center}
\end{figure}


\section{Comparison of the spectra at two light quark masses}\label{sec:comparison}

The principal difference between the spectra presented in Refs.~\cite{Liu:2012,Moir:2013ub} and this work is the light quark mass, corresponding to $M_\pi\sim 400$ MeV in those references and $M_\pi\sim 240$ MeV here.  Figures~\ref{fig:charmonium_comparison}, \ref{fig:Ds_comparison} and \ref{fig:D_comparison} show comparisons of the charmonia, $D_s$ and $D$ spectra at the two light quark masses -- it can be seen that, in general, we observe only a mild light quark mass dependence throughout the entire spectra, with no change in the overall pattern of states. The systematic uncertainties were discussed in Section \ref{sec:calculation_details}.

We note in passing that we achieve a greater statistical precision on the $M_{\pi} \sim 400$ MeV ensemble due to the larger number of time-sources used (see Table \ref{tab:lattice_details}).  In the discussion that follows some notable features in each spectrum are highlighted and in Section \ref{sec:mixing} we investigate the mixing between spin-triplet and spin-singlet open-charm mesons.

\subsection{Charmonium}

\begin{figure}[t]
\includegraphics[width=0.99\textwidth]{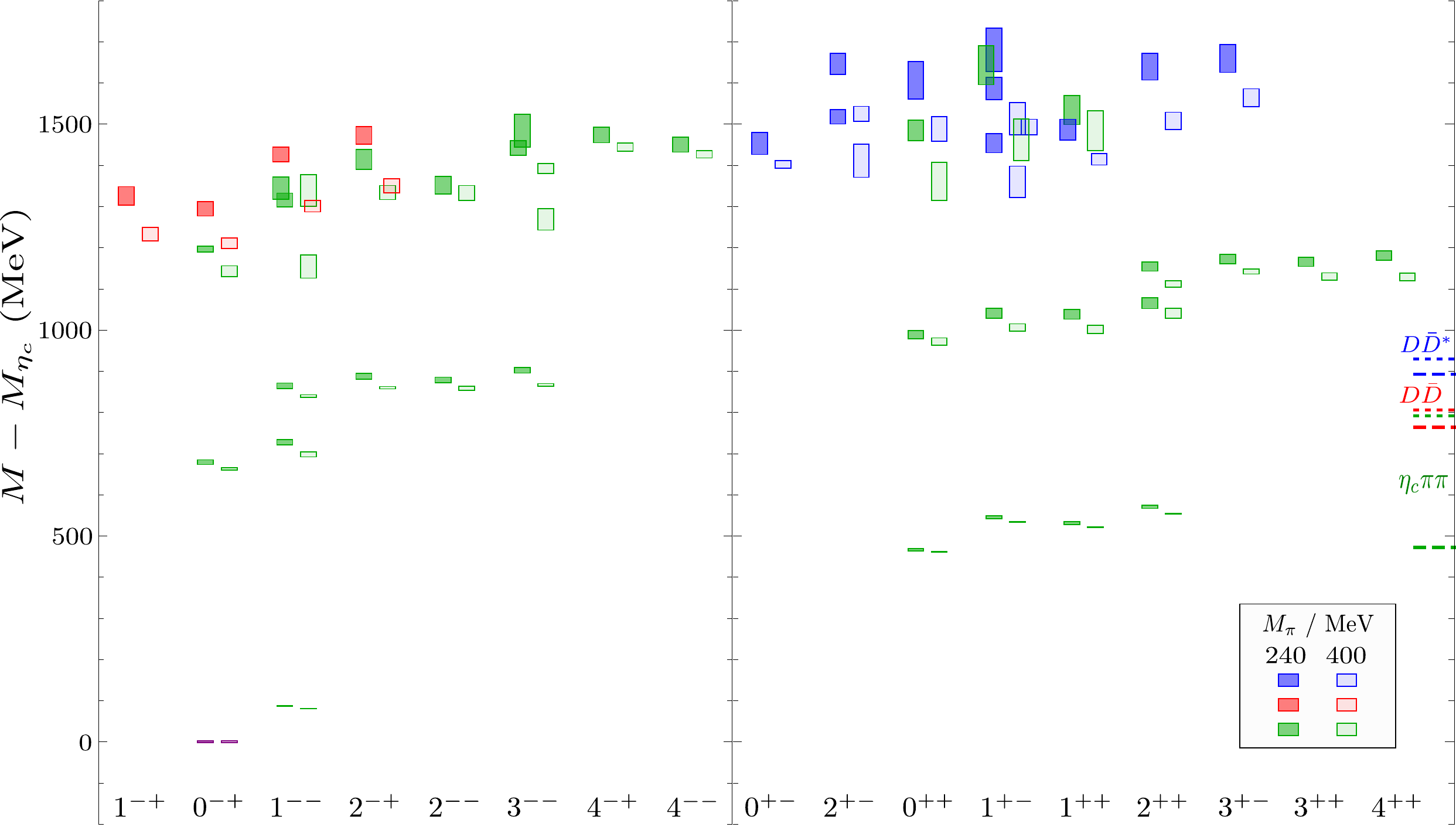}
\caption{Charmonium spectrum, labelled by $J^{PC}$, with $M_\pi\sim 240$ MeV (left column for each $J^{PC}$) compared to the spectrum with $M_\pi\sim 400$ MeV from Ref.~\cite{Liu:2012} (right column for each $J^{PC}$). As in earlier figures, red and blue boxes highlight states identified as constituents of, respectively, the lightest and first-excited supermultiplet of hybrid mesons. Dashed lines show some of the lower thresholds using computed masses for $M_\pi\sim 240$ MeV (coarse dashing) and $M_\pi\sim 400$ MeV (fine dashing): green is $\eta_c \pi \pi$, red is $D \bar{D}$ and blue is $D \bar{D}^*$.}
\label{fig:charmonium_comparison}
\end{figure}

In charmonium the light quark dependence enters through the sea quark content in the dynamical gauge field ensembles. As shown in Figure~\ref{fig:charmonium_comparison}, for the low-lying states the masses are generally consistent between the two ensembles within statistical uncertainties.  An exception is the hyperfine splitting, $M_{J/\psi} - M_{\eta_c}$, where we find a small but statistically significant increase when the light quark mass is decreased.

A second notable feature is that the masses of states higher up in the spectrum are generally larger on the $M_\pi \sim 240$ MeV ensemble.  This is particularly the case for the hybrids, implying a small increase in their mass as $M_\pi$ is reduced; as a consequence the splitting between the hybrids and low-lying conventional mesons increases, albeit in a rather mild fashion. However, it is important to note that at higher energies the statistical uncertainties are larger and neglecting the unstable nature of states may be more important. We emphasise that the overall pattern of hybrid mesons is unaffected by decreasing the light quark mass.

\subsection{$D_s$ mesons}

\begin{figure}[t]
\includegraphics[width=0.99\textwidth]{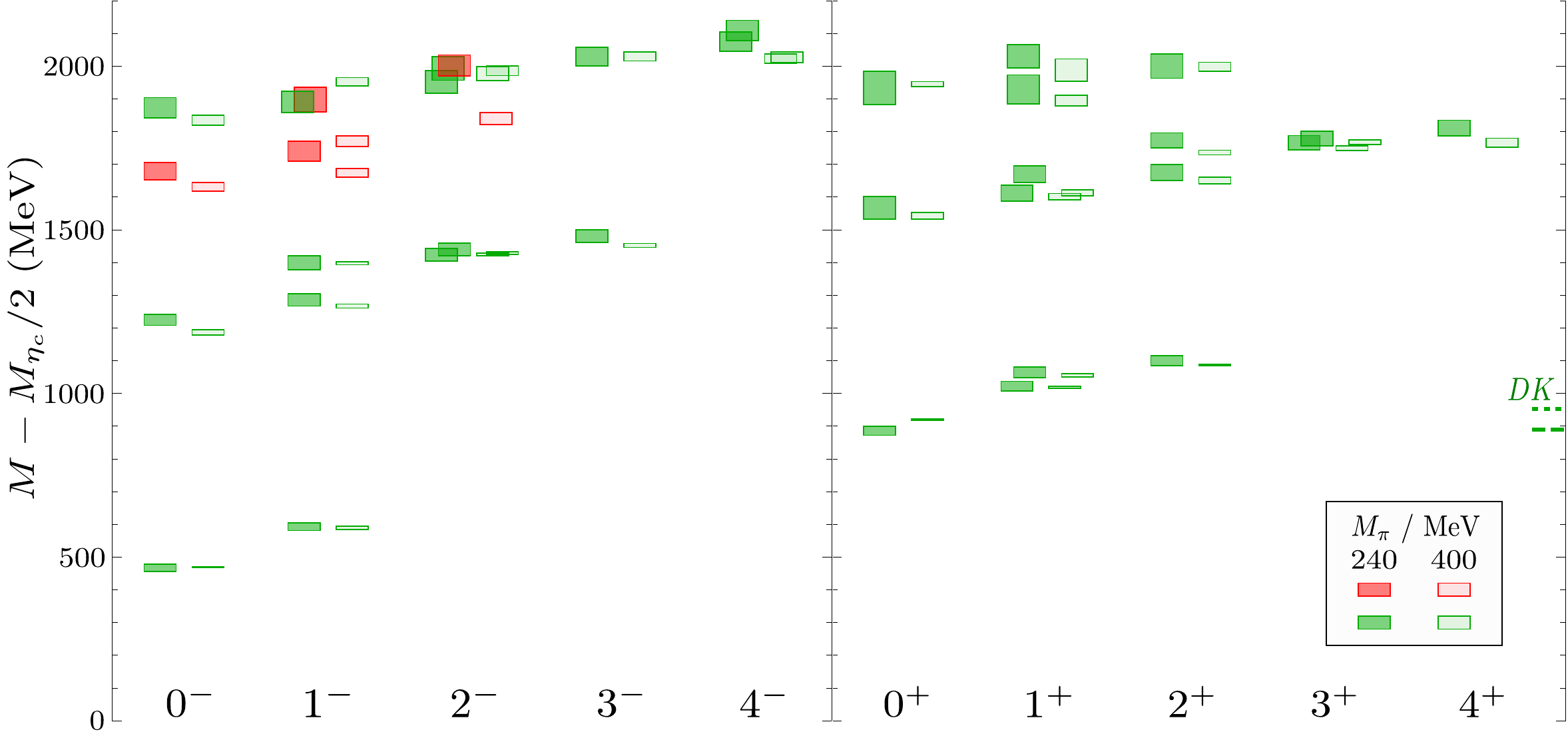}
\caption{As Figure~\ref{fig:charmonium_comparison} but for the $D_{s}$ meson spectrum labelled by $J^P$.}
\label{fig:Ds_comparison}
\end{figure}

\begin{figure}[t]
\includegraphics[width=0.99\textwidth]{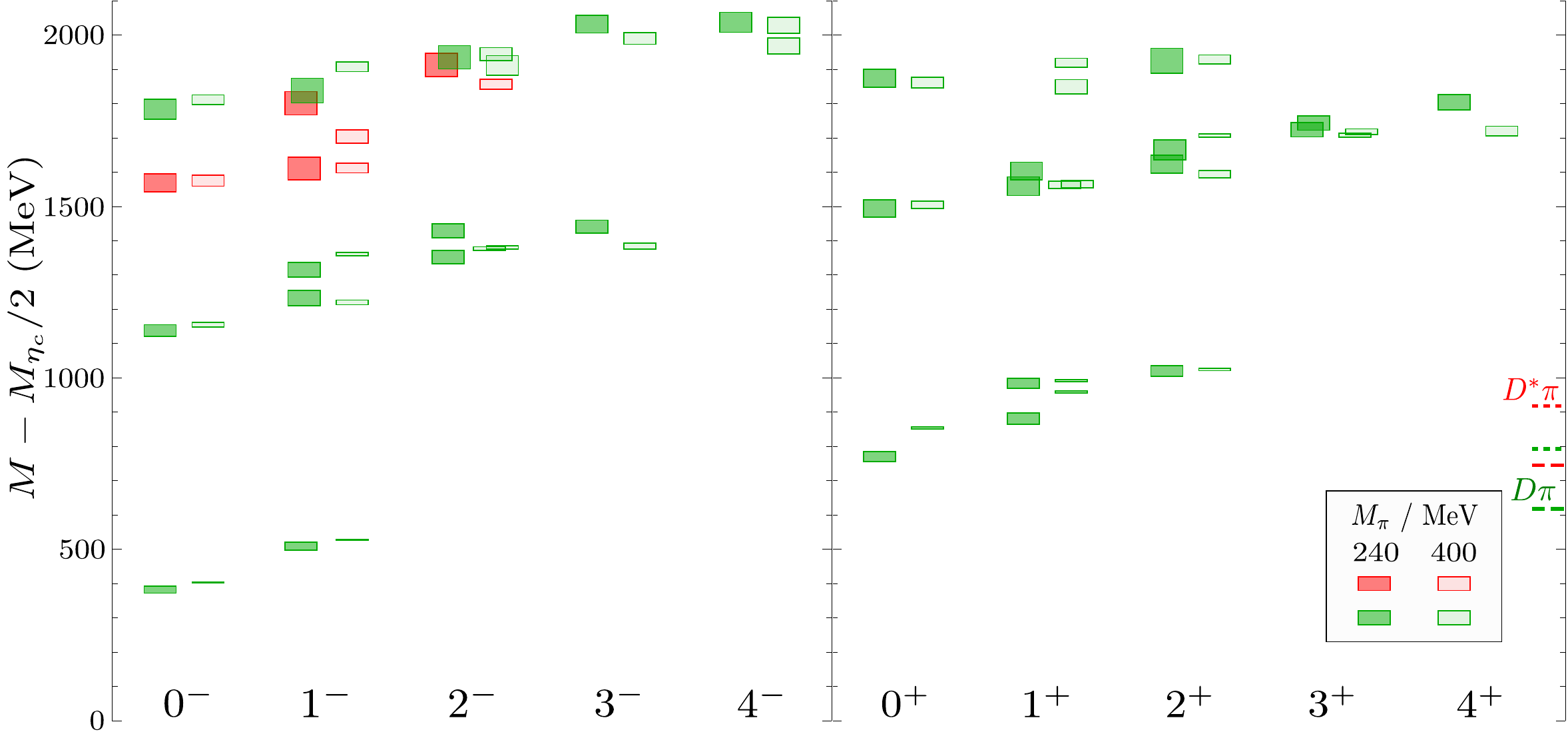}
\caption{As Figure~\ref{fig:charmonium_comparison} but for the $D$ meson spectrum labelled by $J^P$.}
\label{fig:D_comparison}
\end{figure}

As for charmonium, and shown in Fig.\ \ref{fig:Ds_comparison}, only mild dependence on the light quark mass is observed throughout the $D_s$ meson spectrum. The largest change in the low-lying states is for the lightest $0^{+}$ (our candidate for the $D^{*}_{s0}(2317)$). However, this state is expected to be heavily influenced by the nearby $DK$ threshold to which it can couple in $S$ wave, and interestingly, it has decreased just enough to remain below the threshold, in agreement with the experimental situation.

Once again we observe a tendency for the hybrid states, coloured red in Fig.\ \ref{fig:Ds_comparison}, to increase in mass, and hence the splitting between the hybrids and the lowest conventional $D_{s}$ mesons to increase, as $M_\pi$ is reduced.  However, there is no change to their overall pattern.

\subsection{$D$ mesons}

Figure~\ref{fig:D_comparison} shows that, in the $D$ meson spectrum, the light quark mass dependence is also relatively mild and there is no change to the pattern of states.  As expected, the masses generally decrease with decreasing pion mass -- a $D$ meson contains a valence light quark unlike a charmonium or $D_s$ meson.
The most significant differences are observed for the lightest $0^+$ and $1^+$ states and this may be because they are strongly influenced by nearby thresholds that can couple in $S$ wave, namely $D \pi$ and $D^* \pi$ respectively.  However, the mass of the second-lightest $1^+$, which is also in the vicinity of the $D^* \pi$ threshold, does not change significantly.  This may be because the mass difference between the charm quark and the light quark is large enough for the expectations of the heavy-quark limit to be a reasonable guide.  In this limit, one of the $1^+$ states can decay to $D^{*}\pi$ only in $S$ wave, whereas the other can decay to $D^{*}\pi$ only in $D$ wave~\cite{Isgur:1991wq}; the latter would be expected to be influenced less by the position of the $D^* \pi$ threshold.

At higher energies in the spectrum, there are generally only small or statistically insignificant mass shifts while, as for charmonia and $D_{s}$ mesons, there is a general trend for the hybrid mesons to become heavier as the light-quark mass decreases. This change is somewhat less clear in the $D$ meson spectrum because of the opposing trend for mesons to become lighter as the light-quark mass decreases.

\subsection{Mixing of spin-triplet and spin-singlet open-charm mesons}\label{sec:mixing}

As discussed in Section~\ref{sec:spectra:opencharm}, charge conjugation is not a good quantum number for open-charm mesons and, consequently, quark model spin-singlet $(^{1}L_{J=L})$ and spin-triplet $(^{3}L_{J=L})$ states with the same $J=L$ can mix. Quantifying this mixing at different light quark masses can provide an insight into the flavour symmetry breaking. Using a two-state hypothesis and assuming energy-independent mixing we can determine the mixing angle defined in Eq.~(6.1) of~\cite{Moir:2013yfa} from ratios of operator overlaps (interpreted non-relativistically) as described in that reference.

\begin{table}[tb]
\begin{center}
\begin{tabular}{c c c|ccc|c}
&&&&$|\theta|/^{\circ}$&&\\
 & $J^P$ & $M_\pi \, / \, \mathrm{MeV}$ & $\sim(\rho - \rho_2)$ &  $\sim \pi$ & $ \sim \pi_2$ & Heavy-quark limit\\
\hline \hline
c-s &$1^+$         &240 &60.2(0.4) &63.1(0.7) &65.4(0.7) &\multirow{2}{*}{54.7 or 35.3} \\
    &              &400 &60.9(0.6) &64.9(0.2) &66.4(0.4) & \\ \cline{2-7}
    &$2^-$         &240 &56.3(0.9) &60.7(0.8) &63.5(0.9) &\multirow{2}{*}{50.8 or 39.2} \\
    &              &400 &64.9(1.9) &68.7(2.0) &70.9(1.8) & \\ \cline{2-7}
   &$1^-$ (hybrid) &240 &58.9(1.0) &66.2(1.9) &65(2.0) & \\
    &              &400 &59.9(1.7) &67.9(0.9) &67.3(0.9) & \\
\hline \hline
c-l &$1^+$         &240 &52.7(0.9) &61.4(0.4) &67.1(1.0)   &\multirow{2}{*}{54.7 or 35.3} \\
    &              &400 &60.1(0.4) &62.6(0.2) &65.4(0.2) & \\ \cline{2-7}
    &$2^-$         &240 &50.4(0.7) &57.5(0.8) &61.4(0.9) &\multirow{2}{*}{50.8 or 39.2} \\
    &              &400$^{\star}$  &63.3(2.2) &67.8(3.7) &71.1(3.9) & \\ \cline{2-7} 
   &$1^-$ (hybrid) &240 &57.8(1.1) &71.4(2.2) &69.9(2.5) & \\
    &              &400 &59.7(1.1) &68.4(0.8) &67.4(0.9) & \\
\hline \hline
\end{tabular}
\caption{Absolute value of the mixing angles for the lightest pairs of $1^+$, $2^-$ and hybrid $1^-$ states in the charm-strange (c-s) and charm-light (c-l) sectors on the two ensembles.  The mixing angles expected in the heavy-quark limit are also shown.  In the $M_\pi \sim 400$ MeV case highlighted by the $^{\star}$, we have subtracted the angle given in Ref.~\cite{Moir:2013yfa} from $90^{\circ}$ so that the mass ordering of the states is consistent between the two ensembles.}
\label{tab:mixing}
\end{center}
\end{table}

In Table \ref{tab:mixing}, we show the mixing angles for the lightest pairs of $P$-wave ($J^P=1^+$), $D$-wave ($J^{P} = 2^-$) and $ J^{P} = 1^-$ hybrid states extracted using three different operators for the two different ensembles.  The variation between mixing angles determined using the three different operators gives an estimate of the size of the systematic uncertainties as discussed in Ref.~\cite{Moir:2013yfa}.  The $1^+$ mixing angle from the $\rho - \rho_2$ operator in the charm-light sector is closer to the heavy-quark limit value on the $M_\pi \sim 240$ MeV ensemble, but the analogous angle in the charm-strange sector does not differ significantly between the ensembles.  For both charm-light and charm-strange mesons, the $2^{-}$ mixing angle is closer to the heavy-quark limit value for the lighter pion mass whereas the $1^-$ hybrid mixing angle shows no significant difference between the two ensembles.

In all cases on both ensembles, the determined mixing angles lie between the flavour-symmetry limit ($0^{\circ}$ or $90^{\circ}$) and the heavy-quark limit values.  This is expected since the charm quark, although much heavier than the light and strange quarks, is not heavy enough for the heavy-quark limit to apply strictly.

\section{Summary and Outlook}\label{sec:conclusions}

We have presented spectra of excited hidden and open-charm mesons obtained from dynamical lattice QCD calculations with a pion mass of approximately $240$ MeV.  The use of distillation in combination with large variational bases of interpolating operators allows us to extract highly excited mesons, while the spin identification scheme has allowed a robust identification of the $J^{P(C)}$ of states as high as spin four, including states with exotic quantum numbers.  
The majority of mesons we extract can be interpreted in terms of the $n^{2S+1}L_{J}$ pattern expected from quark potential models.  However, excess states, with both exotic and non-exotic quantum numbers, that do not fit this pattern are also determined.  By examining the operator overlaps we identify these as hybrid mesons, i.e. having excited gluonic degrees of freedom.  The supermultiplets of hybrid mesons follow a pattern consistent with a quark-antiquark combination in $S$ or $P$-wave coupled to a $1^{+-}$ gluonic excitation.
The pattern and energy scale of hybrids are the same as that found in the light meson and baryon sectors~\cite{Dudek:2010,Dudek:2011b,Dudek:2012,Edwards:2012fx,Dudek:2013yja}, studies of charmed baryons~\cite{Padmanath:2013zfa,Padmanath:2015jea} and in our earlier work on charmonia and open-charm mesons~\cite{Liu:2012,Moir:2013ub}.

Comparing the spectra to those from a similar lattice calculation with a pion mass of approximately $400$ MeV, we find that the overall qualitative features 
are the same and, even in the case of charm mesons with a valence light quark, we find only small quantitative differences. 
The hybrid mesons appear to show a mild increase in mass as the pion mass is decreased but the pattern of states and supermultiplet structure is unchanged.

We also compared the spin-singlet -- spin-triplet mixing angles for the lightest pairs of charm-strange and charm-light $P$-wave $(J^{P} = 1^{+})$, $D$-wave $(J^{P} = 2^{-})$ and hybrid $(J^{P} = 1^{-})$ states between the two lattice ensembles.  Using a non-relativistic interpretation of operator overlaps, our results suggest that the mixing angles for the charm-light $1^{+}$ and the charm-light and charm-strange $2^{-}$ states become closer to those expected in the heavy-quark limit as the pion mass is reduced.  Conversely, we find no significant difference in the hybrid $1^{-}$ mixing angles between the two ensembles.

As discussed earlier, a limitation of these calculations is that we have not accounted for the unstable nature of states above threshold.  
This issue has already been addressed for a variety of mesons appearing as bound states and resonances in coupled-channel $D\pi$, $D\eta$ and $D_{s}\bar{K}$ scattering~\cite{Moir:2016srx}.  The work presented here lays the foundation for extending these scattering calculations to pion masses 
closer to the physical value, as well as to other scattering channels involving hidden and open-charm mesons.

\bigskip

\begin{acknowledgments}
We thank our colleagues in the Hadron Spectrum Collaboration. 
GC is supported by the Cambridge European Trust, the U.K. Science and Technology Facilities Council (STFC), and St John's College, Cambridge. 
COH acknowledges support from the School of Mathematics at Trinity College Dublin.
GM acknowledges support the Herchel Smith Fund at the University of Cambridge.
SMR acknowledges support from Science Foundation Ireland [RFP-PHY-3201].
CET acknowledges support from the STFC [grant ST/L000385/1].
DT is supported by the Irish Research Council Government of Ireland Postgraduate Scholarship Scheme [grant GOIPG/2014/65]. 

This work used the DiRAC Complexity system, operated by the University of Leicester IT Services, which forms part of the STFC DiRAC HPC Facility (www.dirac.ac.uk). This equipment is funded by BIS National E-Infrastructure capital grant ST/K000373/1 and STFC DiRAC Operations grant ST/K0003259/1. DiRAC is part of the National E-Infrastructure.
This work also used the Wilkes GPU cluster at the University of Cambridge High Performance Computing Service (http://www.hpc.cam.ac.uk/), provided by Dell Inc., NVIDIA and Mellanox, and part funded by STFC with industrial sponsorship from Rolls Royce and Mitsubishi Heavy Industries.
Computations were also performed at Jefferson Laboratory under the USQCD Initiative and the LQCD ARRA project and on the Lonsdale cluster maintained by the Trinity Centre for High Performance Computing (TCHPC) funded through grants from Science Foundation Ireland (SFI).

The software codes {\tt Chroma}~\cite{Edwards:2004sx}, {\tt QUDA}~\cite{Clark:2009wm,Babich:2010mu}, {\tt QPhiX}~\cite{ISC13Phi}, and {\tt QOPQDP}~\cite{Osborn:2010mb,Babich:2010qb} were used to compute the propagators required for this project.
This research was supported in part under an ALCC award, and used resources of the Oak Ridge Leadership Computing Facility at the Oak Ridge National Laboratory, which is supported by the Office of Science of the U.S. Department of Energy under Contract No. DE-AC05-00OR22725. This research is also part of the Blue Waters sustained-petascale computing project, which is supported by the National Science Foundation (awards OCI-0725070 and ACI-1238993) and the state of Illinois. Blue Waters is a joint effort of the University of Illinois at Urbana-Champaign and its National Center for Supercomputing Applications. This work is also part of the PRAC ``Lattice QCD on Blue Waters''. This research used resources of the National Energy Research Scientific Computing Center (NERSC), a DOE Office of Science User Facility supported by the Office of Science of the U.S. Department of Energy under Contract No. DEAC02-05CH11231. The authors acknowledge the Texas Advanced Computing Center (TACC) at The University of Texas at Austin for providing HPC resources that have contributed to the research results reported within this paper.

Gauge configurations were generated using resources awarded from the U.S. Department of Energy INCITE program at the Oak Ridge Leadership Computing Facility, the NERSC, the NSF Teragrid at the TACC and the Pittsburgh Supercomputer Center, as well as at Jefferson Lab.
\end{acknowledgments}

\appendix
\section{Tables of Results}\label{app:tables}

In Tables \ref{tab:charmonium_summary}, \ref{tab:Ds_summary} and \ref{tab:D_summary} we present numerical values for, respectively, the charmonium, $D_{s}$ and $D$ meson masses obtained for $M_\pi \sim 240$~MeV. 
Masses are given in MeV with either the mass of the $\eta_{c}$ subtracted (charmonium) or half the mass of the $\eta_{c}$ subtracted (open-charm mesons)
in order to minimise the systematic uncertainty in tuning the charm quark mass. 
In all cases the quoted error corresponds to the (one-sigma) statistical uncertainty.  As discussed earlier, above the lowest multi-hadron threshold in each channel states can decay strongly into lighter hadrons and, aside from any other systematic uncertainties, we only expect the masses to be correct up to around the width of the state~\cite{Dudek:2010}.

\begin{table}[h!]
\begin{center}
\begin{tabular}{|c|cccccc|}
\hline
$J^{PC}$ &  &  & $M-M_{\eta_c}$ &\hspace{-0.7cm} $(MeV)$  &  & \\
\hline
$0^{-+}$ & 0 & 679(6) & 1197(7) & 1295(18)&  &  \\
$1^{--}$ & 88(1) & 728(7) & 865(7) & 1316(17)& 1345(27) & 1427(17) \\
$2^{--}$ & 879(7) & 1352(21) &  & &  &  \\
$2^{-+}$ & 888(7) & 1414(24) & 1472(21) &  &  &  \\
$3^{--}$ & 902(6) & 1442(18) & 1484(40) & &  &  \\
$4^{-+}$ & 1474(19)  &  &  & &  &  \\
$4^{--}$ & 1450(18) &  &  & &  &  \\
\hline
$0^{++}$ & 466(3)  & 989(10) & 1485(25) & 1607(46)&  &  \\
$1^{++}$ & 531(4) & 1038(12) & 1486(25) & 1534(35)&  &  \\
$1^{+-}$ & 545(4) & 1041(12) & 1454(23) & 1587(27) & 1643(47) & 1681(53)   \\
$2^{++}$ & 571(4) & 1065(13) & 1154(11) & 1173(11) & 1639(32) &  \\
$3^{++}$ & 1166(11) &  &  & &  &  \\
$3^{+-}$ & 1173(11) & 1660(34) &  & &  &  \\
$4^{++}$ & 1181(12) &  &  & &  &  \\
\hline
$1^{-+}$ & 1326(23) &  &  & &  &  \\
$0^{+-}$ & 1453(27) &  &  & &  &  \\
$2^{+-}$ & 1518(18) & 1647(26) &  & &  &  \\
\hline
\end{tabular}
\caption{Summary of the charmonium spectrum presented in Figure \ref{fig:charmonium_spectrum}.  Masses are shown with $M_{\eta_{c}}$ subtracted.  Quoted uncertainties are statistical only.}
\label{tab:charmonium_summary}
\end{center}
\end{table}

\begin{table}[h!]
\begin{center}
\begin{tabular}{|c|cccccc|}
\hline
$J^P$ &  &  & $M-M_{\eta_c}/2$& \hspace{-0.7cm}$(MeV)$  &  &   \\
\hline
$0^-$ & 467(11) & 1225(17) & 1679(27) & 1873(31) & & \\
$1^-$ & 593(12) & 1286(12) & 1399(21) & 1740(30) & 1891(33) & 1898(38)\\
$2^-$ & 1424(19) & 1440(20) & 1952(35) & 1993(36) & 2002(32) & \\
$3^-$ & 1481(19) & 2029(28) &  &  &  & \\
$4^-$ & 2075(29) & 2109(31) &  &  &  & \\
\hline
$0^+$ & 886(14)  & 1567(35) & 1934(51) & &  & \\
$1^+$ & 1022(15) & 1064(16) & 1612(25) & 1670(26) & 1929(44) & 2030(35) \\
$2^+$ & 1100(15) & 1675(24) & 1773(23) & 2000(37) & & \\
$3^+$ & 1766(22) & 1779(22) &  &  &  & \\
$4^+$ & 1811(24) &  &  &  &  & \\
\hline
\end{tabular}
\caption{Summary of the $D_{s}$ meson spectrum presented in Figure \ref{fig:Ds_spectrum}.  Masses are shown with $M_{\eta_{c}} / 2$ subtracted.  Quoted uncertainties are statistical only.}
\label{tab:Ds_summary}
\end{center}
\end{table}

\clearpage

\begin{table}[h!]
\begin{center}
\begin{tabular}{|c|cccccc|}
\hline
$J^P$ &  &  & $M-M_{\eta_c}/2$ &\hspace{-0.7cm}$(MeV)$  &  &   \\
\hline
$0^-$ & 382(10) & 1138(17) & 1569(26) & 1783(29) & 2176(37)  &\\
$1^-$ & 509(11) & 1233(22) & 1315(21) & 1610(33) & 1801(34) & 1838(36)\\
$2^-$ & 1352(19) & 1429(20) & 1912(34) & 1935(34) & & \\
$3^-$ & 1441(19) & 2032(26) &  &  &  & \\
$4^-$ & 2037(29) & &  &  &  & \\
\hline
$0^+$ & 770(15) & 1494(25) & 2201(45) & 1874(26) &  & \\
$1^+$ & 881(17) & 984(14) & 1559(27) & 1603(26) & &\\
$2^+$ & 1020(16) & 1623(26) & 1665(29) & 1925(36) & & \\
$3^+$ & 1724(21) & 1743(21) &  &  &  & \\
$4^+$ & 1804(22) &  &  &  &  & \\
\hline
\end{tabular}
\caption{Summary of the $D$ meson spectrum presented in Figure \ref{fig:D_spectrum}.  Masses are shown with $M_{\eta_{c}} / 2$ subtracted.  Quoted uncertainties are statistical only.}
\label{tab:D_summary}
\end{center}
\end{table}


\bibliography{charm_860_paper}

\providecommand{\href}[2]{#2}\begingroup\raggedright\begin{thebibliography}{10}

\bibitem{PDG2015}
{\bf Particle Data Group} Collaboration, K.~A. Olive et~al., {\it {Review of
  Particle Physics}},  {\em Chin. Phys.} {\bf C38} (2014) 090001 and 2015
  update.

\bibitem{Brambilla:2010cs}
N.~Brambilla et~al., {\it {Heavy quarkonium: progress, puzzles, and
  opportunities}},  {\em Eur. Phys. J.} {\bf C71} (2011) 1534,
  [\href{http://arxiv.org/abs/1010.5827}{{\tt arXiv:1010.5827}}].

\bibitem{Brambilla:2014jmp}
N.~Brambilla et~al., {\it {QCD and Strongly Coupled Gauge Theories: Challenges
  and Perspectives}},  {\em Eur. Phys. J.} {\bf C74} (2014), no.~10 2981,
  [\href{http://arxiv.org/abs/1404.3723}{{\tt arXiv:1404.3723}}].

\bibitem{Olsen:2015zcy}
S.~L. Olsen, {\it {XYZ Meson Spectroscopy}},  in {\em {Proceedings, 53rd
  International Winter Meeting on Nuclear Physics (Bormio 2015): Bormio, Italy,
  January 26-30, 2015}}, 2015.
\newblock \href{http://arxiv.org/abs/1511.01589}{{\tt arXiv:1511.01589}}.

\bibitem{Swanson:2015wgq}
E.~S. Swanson, {\it {XYZ States: Theory Overview}},  {\em AIP Conf. Proc.} {\bf
  1735} (2016) 020013, [\href{http://arxiv.org/abs/1512.04853}{{\tt
  arXiv:1512.04853}}].

\bibitem{Prencipe:2015kva}
E.~Prencipe, {\it {Hadrons with c-s content: past, present and future}},  in
  {\em {Proceedings, 53rd International Winter Meeting on Nuclear Physics
  (Bormio 2015): Bormio, Italy, January 26-30, 2015}}, 2015.
\newblock \href{http://arxiv.org/abs/1510.03053}{{\tt arXiv:1510.03053}}.

\bibitem{Namekawa:2011wt}
{\bf PACS-CS Collaboration} Collaboration, Y.~Namekawa et~al., {\it {Charm
  quark system at the physical point of 2+1 flavor lattice QCD}},  {\em
  Phys.Rev.} {\bf D84} (2011) 074505,
  [\href{http://arxiv.org/abs/1104.4600}{{\tt arXiv:1104.4600}}].

\bibitem{McNeile:2012qf}
C.~McNeile, C.~Davies, E.~Follana, K.~Hornbostel, and G.~Lepage, {\it {Heavy
  meson masses and decay constants from relativistic heavy quarks in full
  lattice QCD}},  {\em Phys.Rev.} {\bf D86} (2012) 074503,
  [\href{http://arxiv.org/abs/1207.0994}{{\tt arXiv:1207.0994}}].

\bibitem{Dowdall:2012ab}
R.~Dowdall, C.~Davies, T.~Hammant, and R.~Horgan, {\it {Precise heavy-light
  meson masses and hyperfine splittings from lattice QCD including charm quarks
  in the sea}},  {\em Phys.Rev.} {\bf D86} (2012) 094510,
  [\href{http://arxiv.org/abs/1207.5149}{{\tt arXiv:1207.5149}}].

\bibitem{Donald:2012ga}
G.~C. Donald, C.~T.~H. Davies, R.~J. Dowdall, E.~Follana, K.~Hornbostel,
  J.~Koponen, G.~P. Lepage, and C.~McNeile, {\it {Precision tests of the
  $J/{\psi}$ from full lattice QCD: mass, leptonic width and radiative decay
  rate to ${\eta}_c$}},  {\em Phys. Rev.} {\bf D86} (2012) 094501,
  [\href{http://arxiv.org/abs/1208.2855}{{\tt arXiv:1208.2855}}].

\bibitem{Galloway:2014tta}
B.~A. Galloway, P.~Knecht, J.~Koponen, C.~T.~H. Davies, and G.~P. Lepage, {\it
  {Radial and orbital excitation energies of charmonium}},  {\em PoS} {\bf
  LATTICE2014} (2014) 092, [\href{http://arxiv.org/abs/1411.1318}{{\tt
  arXiv:1411.1318}}].

\bibitem{Dudek:2007}
J.~J. Dudek, R.~G. Edwards, N.~Mathur, and D.~G. Richards, {\it {Charmonium
  excited state spectrum in lattice QCD}},  {\em Phys. Rev.} {\bf D77} (2008)
  034501, [\href{http://arxiv.org/abs/0707.4162}{{\tt arXiv:0707.4162}}].

\bibitem{Bali:2011rd}
G.~S. Bali, S.~Collins, and C.~Ehmann, {\it {Charmonium spectroscopy and mixing
  with light quark and open charm states from $n_F$=2 lattice QCD}},  {\em
  Phys. Rev.} {\bf D84} (2011) 094506,
  [\href{http://arxiv.org/abs/1110.2381}{{\tt arXiv:1110.2381}}].

\bibitem{Bali:2011dc}
G.~Bali et~al., {\it {Spectra of heavy-light and heavy-heavy mesons containing
  charm quarks, including higher spin states for $N_f=2+ 1$}},  {\em PoS} {\bf
  LATTICE2011} (2011) 135, [\href{http://arxiv.org/abs/1108.6147}{{\tt
  arXiv:1108.6147}}].

\bibitem{Mohler:2011ke}
D.~Mohler and R.~M. Woloshyn, {\it {$D$ and $D_s$ meson spectroscopy}},  {\em
  Phys. Rev.} {\bf D84} (2011) 054505,
  [\href{http://arxiv.org/abs/1103.5506}{{\tt arXiv:1103.5506}}].

\bibitem{Bali:2015lka}
P.~P\'{e}rez-Rubio, S.~Collins, and G.~S. Bali, {\it {Charmed baryon
  spectroscopy and light flavor symmetry from lattice QCD}},  {\em Phys. Rev.}
  {\bf D92} (2015), no.~3 034504, [\href{http://arxiv.org/abs/1503.08440}{{\tt
  arXiv:1503.08440}}].

\bibitem{Kalinowski:2015bwa}
M.~Kalinowski and M.~Wagner, {\it {Masses of $D$ mesons, $D_s$ mesons and
  charmonium states from twisted mass lattice QCD}},  {\em Phys. Rev.} {\bf
  D92} (2015), no.~9 094508, [\href{http://arxiv.org/abs/1509.02396}{{\tt
  arXiv:1509.02396}}].

\bibitem{Cichy:2016bci}
K.~Cichy, M.~Kalinowski, and M.~Wagner, {\it {The continuum limit of the $D$
  meson, $D_s$ meson and charmonium spectrum from $N_f=2+1+1$ twisted mass
  lattice QCD}},  \href{http://arxiv.org/abs/1603.06467}{{\tt
  arXiv:1603.06467}}.

\bibitem{Berwein:2015vca}
M.~Berwein, N.~Brambilla, J.~Tarrús~Castellà, and A.~Vairo, {\it {Quarkonium
  Hybrids with Nonrelativistic Effective Field Theories}},  {\em Phys. Rev.}
  {\bf D92} (2015), no.~11 114019, [\href{http://arxiv.org/abs/1510.04299}{{\tt
  arXiv:1510.04299}}].

\bibitem{Guo:2012tg}
F.-K. Guo and U.-G. Meissner, {\it {Light quark mass dependence in heavy
  quarkonium physics}},  {\em Phys. Rev. Lett.} {\bf 109} (2012) 062001,
  [\href{http://arxiv.org/abs/1203.1116}{{\tt arXiv:1203.1116}}].

\bibitem{Dudek:2010}
J.~J. Dudek, R.~G. Edwards, M.~J. Peardon, D.~G. Richards, and C.~E. Thomas,
  {\it {Toward the excited meson spectrum of dynamical QCD}},  {\em Phys. Rev.}
  {\bf D82} (2010) 034508, [\href{http://arxiv.org/abs/1004.4930}{{\tt
  arXiv:1004.4930}}].

\bibitem{Liu:2012}
L.~Liu et~al., {\it {Excited and exotic charmonium spectroscopy from lattice
  QCD}},  {\em JHEP} {\bf 07} (2012) 126,
  [\href{http://arxiv.org/abs/1204.5425}{{\tt arXiv:1204.5425}}].

\bibitem{Moir:2013ub}
G.~Moir, M.~Peardon, S.~M. Ryan, C.~E. Thomas, and L.~Liu, {\it {Excited
  spectroscopy of charmed mesons from lattice QCD}},  {\em JHEP} {\bf 05}
  (2013) 021, [\href{http://arxiv.org/abs/1301.7670}{{\tt arXiv:1301.7670}}].

\bibitem{Moir:2016srx}
G.~Moir, M.~Peardon, S.~M. Ryan, C.~E. Thomas, and D.~J. Wilson, {\it
  {Coupled-Channel $D\pi$, $D\eta$ and $D_{s}\bar{K}$ Scattering from Lattice
  QCD}},  {\em JHEP} {\bf 10} (2016) 011,
  [\href{http://arxiv.org/abs/1607.07093}{{\tt arXiv:1607.07093}}].

\bibitem{Ozaki:2012ce}
S.~Ozaki and S.~Sasaki, {\it {Lüscher’s finite size method with twisted
  boundary conditions: An application to the $J/\psi$-$\phi$ system to search
  for a narrow resonance}},  {\em Phys. Rev.} {\bf D87} (2013), no.~1 014506,
  [\href{http://arxiv.org/abs/1211.5512}{{\tt arXiv:1211.5512}}].

\bibitem{Mohler:2012na}
D.~Mohler, S.~Prelovsek, and R.~M. Woloshyn, {\it {$D \pi$ scattering and $D$
  meson resonances from lattice QCD}},  {\em Phys. Rev.} {\bf D87} (2013),
  no.~3 034501, [\href{http://arxiv.org/abs/1208.4059}{{\tt arXiv:1208.4059}}].

\bibitem{Liu:2012zya}
L.~Liu, K.~Orginos, F.-K. Guo, C.~Hanhart, and U.-G. Meissner, {\it
  {Interactions of charmed mesons with light pseudoscalar mesons from lattice
  QCD and implications on the nature of the $D_{s0}^*(2317)$}},  {\em Phys.
  Rev.} {\bf D87} (2013), no.~1 014508,
  [\href{http://arxiv.org/abs/1208.4535}{{\tt arXiv:1208.4535}}].

\bibitem{Prelovsek:2013cra}
S.~Prelovsek and L.~Leskovec, {\it {Evidence for X(3872) from DD* scattering on
  the lattice}},  {\em Phys. Rev. Lett.} {\bf 111} (2013) 192001,
  [\href{http://arxiv.org/abs/1307.5172}{{\tt arXiv:1307.5172}}].

\bibitem{Prelovsek:2013xba}
S.~Prelovsek and L.~Leskovec, {\it {Search for $Z^{+}_{c}$(3900) in the
  $1^{+-}$ Channel on the Lattice}},  {\em Phys. Lett.} {\bf B727} (2013)
  172--176, [\href{http://arxiv.org/abs/1308.2097}{{\tt arXiv:1308.2097}}].

\bibitem{Mohler:2013rwa}
D.~Mohler, C.~B. Lang, L.~Leskovec, S.~Prelovsek, and R.~M. Woloshyn, {\it
  {$D_{s0}^*(2317)$ Meson and $D$-Meson-Kaon Scattering from Lattice QCD}},
  {\em Phys. Rev. Lett.} {\bf 111} (2013), no.~22 222001,
  [\href{http://arxiv.org/abs/1308.3175}{{\tt arXiv:1308.3175}}].

\bibitem{Chen:2014afa}
Y.~Chen et~al., {\it {Low-energy scattering of the $(D\bar{D}^*)^\pm$ system
  and the resonance-like structure $Z_c(3900)$}},  {\em Phys. Rev.} {\bf D89}
  (2014), no.~9 094506, [\href{http://arxiv.org/abs/1403.1318}{{\tt
  arXiv:1403.1318}}].

\bibitem{Lang:2014yfa}
C.~B. Lang, L.~Leskovec, D.~Mohler, S.~Prelovsek, and R.~M. Woloshyn, {\it {Ds
  mesons with DK and D*K scattering near threshold}},  {\em Phys. Rev.} {\bf
  D90} (2014), no.~3 034510, [\href{http://arxiv.org/abs/1403.8103}{{\tt
  arXiv:1403.8103}}].

\bibitem{Prelovsek:2014swa}
S.~Prelovsek, C.~B. Lang, L.~Leskovec, and D.~Mohler, {\it {Study of the
  $Z_c^+$ channel using lattice QCD}},  {\em Phys. Rev.} {\bf D91} (2015),
  no.~1 014504, [\href{http://arxiv.org/abs/1405.7623}{{\tt arXiv:1405.7623}}].

\bibitem{Padmanath:2015era}
M.~Padmanath, C.~B. Lang, and S.~Prelovsek, {\it {X(3872) and Y(4140) using
  diquark-antidiquark operators with lattice QCD}},  {\em Phys. Rev.} {\bf D92}
  (2015), no.~3 034501, [\href{http://arxiv.org/abs/1503.03257}{{\tt
  arXiv:1503.03257}}].

\bibitem{Lang:2015sba}
C.~B. Lang, L.~Leskovec, D.~Mohler, and S.~Prelovsek, {\it {Vector and scalar
  charmonium resonances with lattice QCD}},  {\em JHEP} {\bf 09} (2015) 089,
  [\href{http://arxiv.org/abs/1503.05363}{{\tt arXiv:1503.05363}}].

\bibitem{Chen:2015jwa}
{\bf CLQCD} Collaboration, Y.~Chen et~al., {\it {Low-energy Scattering of
  $(D^{*}\bar{D}^{*})^\pm$ System and the Resonance-like Structure
  $Z_c(4025)$}},  {\em Phys. Rev.} {\bf D92} (2015), no.~5 054507,
  [\href{http://arxiv.org/abs/1503.02371}{{\tt arXiv:1503.02371}}].

\bibitem{Chen:2016lkl}
{\bf CLQCD} Collaboration, T.~Chen et~al., {\it {A Lattice Study of $(\bar{D}_1
  D^{*})^\pm$ Near-threshold Scattering}},  {\em Phys. Rev.} {\bf D93} (2016),
  no.~11 114501, [\href{http://arxiv.org/abs/1602.00200}{{\tt
  arXiv:1602.00200}}].

\bibitem{Ikeda:2016zwx}
Y.~Ikeda, S.~Aoki, T.~Doi, S.~Gongyo, T.~Hatsuda, T.~Inoue, T.~Iritani,
  N.~Ishii, K.~Murano, and K.~Sasaki, {\it {Fate of the Tetraquark Candidate
  Zc(3900) in Lattice QCD}},  \href{http://arxiv.org/abs/1602.03465}{{\tt
  arXiv:1602.03465}}.

\bibitem{Bicudo:2012qt}
{\bf European Twisted Mass} Collaboration, P.~Bicudo and M.~Wagner, {\it
  {Lattice QCD signal for a bottom-bottom tetraquark}},  {\em Phys. Rev.} {\bf
  D87} (2013), no.~11 114511, [\href{http://arxiv.org/abs/1209.6274}{{\tt
  arXiv:1209.6274}}].

\bibitem{Brown:2012tm}
Z.~S. Brown and K.~Orginos, {\it {Tetraquark bound states in the heavy-light
  heavy-light system}},  {\em Phys. Rev.} {\bf D86} (2012) 114506,
  [\href{http://arxiv.org/abs/1210.1953}{{\tt arXiv:1210.1953}}].

\bibitem{Ikeda:2013vwa}
Y.~Ikeda, B.~Charron, S.~Aoki, T.~Doi, T.~Hatsuda, T.~Inoue, N.~Ishii,
  K.~Murano, H.~Nemura, and K.~Sasaki, {\it {Charmed tetraquarks $T_{cc}$ and
  $T_{cs}$ from dynamical lattice QCD simulations}},  {\em Phys. Lett.} {\bf
  B729} (2014) 85--90, [\href{http://arxiv.org/abs/1311.6214}{{\tt
  arXiv:1311.6214}}].

\bibitem{Bicudo:2015vta}
P.~Bicudo, K.~Cichy, A.~Peters, B.~Wagenbach, and M.~Wagner, {\it {Evidence for
  the existence of $u d \bar{b} \bar{b}$ and the non-existence of $s s \bar{b}
  \bar{b}$ and $c c \bar{b} \bar{b}$ tetraquarks from lattice QCD}},  {\em
  Phys. Rev.} {\bf D92} (2015), no.~1 014507,
  [\href{http://arxiv.org/abs/1505.00613}{{\tt arXiv:1505.00613}}].

\bibitem{Bicudo:2015kna}
P.~Bicudo, K.~Cichy, A.~Peters, and M.~Wagner, {\it {BB interactions with
  static bottom quarks from Lattice QCD}},  {\em Phys. Rev.} {\bf D93} (2016),
  no.~3 034501, [\href{http://arxiv.org/abs/1510.03441}{{\tt
  arXiv:1510.03441}}].

\bibitem{Francis:2016hui}
A.~Francis, R.~J. Hudspith, R.~Lewis, and K.~Maltman, {\it {Doubly bottom
  strong-interaction stable tetraquarks from lattice QCD}},
  \href{http://arxiv.org/abs/1607.05214}{{\tt arXiv:1607.05214}}.

\bibitem{Alberti:2016dru}
M.~Alberti, G.~S. Bali, S.~Collins, F.~Knechtli, G.~Moir, and W.~Söldner, {\it
  {Hadro-quarkonium from Lattice QCD}},
  \href{http://arxiv.org/abs/1608.06537}{{\tt arXiv:1608.06537}}.

\bibitem{Peters:2016isf}
A.~Peters, P.~Bicudo, L.~Leskovec, S.~Meinel, and M.~Wagner, {\it {Lattice QCD
  study of heavy-heavy-light-light tetraquark candidates}},  in {\em
  {Proceedings, 34th International Symposium on Lattice Field Theory (Lattice
  2016): Southampton, UK, July 24-30, 2016}}, 2016.
\newblock \href{http://arxiv.org/abs/1609.00181}{{\tt arXiv:1609.00181}}.

\bibitem{Bicudo:2016jwl}
P.~Bicudo, J.~Scheunert, and M.~Wagner, {\it {Including heavy spin effects in a
  lattice QCD study of static-static-light-light tetraquarks}},  in {\em
  {Proceedings, 34th International Symposium on Lattice Field Theory (Lattice
  2016): Southampton, UK, July 24-30, 2016}}, 2016.
\newblock \href{http://arxiv.org/abs/1609.00548}{{\tt arXiv:1609.00548}}.

\bibitem{Peardon:2009}
{\bf Hadron Spectrum} Collaboration, M.~Peardon et~al., {\it {A novel
  quark-field creation operator construction for hadronic physics in lattice
  QCD}},  {\em Phys. Rev.} {\bf D80} (2009) 054506,
  [\href{http://arxiv.org/abs/0905.2160}{{\tt arXiv:0905.2160}}].

\bibitem{Morningstar:2003}
C.~Morningstar and M.~J. Peardon, {\it {Analytic smearing of SU(3) link
  variables in lattice QCD}},  {\em Phys. Rev.} {\bf D69} (2004) 054501,
  [\href{http://arxiv.org/abs/hep-lat/0311018}{{\tt hep-lat/0311018}}].

\bibitem{Edwards:2008}
R.~G. Edwards, B.~Jo\'o, and H.-W. Lin, {\it {Tuning for Three-flavors of
  Anisotropic Clover Fermions with Stout-link Smearing}},  {\em Phys. Rev.}
  {\bf D78} (2008) 054501, [\href{http://arxiv.org/abs/0803.3960}{{\tt
  arXiv:0803.3960}}].

\bibitem{Lin:2009}
{\bf Hadron Spectrum} Collaboration, H.-W. Lin et~al., {\it {First results from
  2+1 dynamical quark flavors on an anisotropic lattice: Light-hadron
  spectroscopy and setting the strange-quark mass}},  {\em Phys. Rev.} {\bf
  D79} (2009) 034502, [\href{http://arxiv.org/abs/0810.3588}{{\tt
  arXiv:0810.3588}}].

\bibitem{Wilson:2015dqa}
D.~J. Wilson, R.~A. Brice\~no, J.~J. Dudek, R.~G. Edwards, and C.~E. Thomas,
  {\it {Coupled $\pi\pi, K\bar{K}$ scattering in $P$-wave and the $\rho$
  resonance from lattice QCD}},  {\em Phys. Rev.} {\bf D92} (2015), no.~9
  094502, [\href{http://arxiv.org/abs/1507.02599}{{\tt arXiv:1507.02599}}].

\bibitem{Johnson:1982yq}
R.~C. Johnson, {\it {Angular momentum on a lattice}},  {\em Phys. Lett.} {\bf
  B114} (1982) 147.

\bibitem{Michael:1985}
C.~Michael, {\it {Adjoint Sources in Lattice Gauge Theory}},  {\em Nucl. Phys.}
  {\bf B259} (1985) 58.

\bibitem{Luscher:1990}
M.~L\"uscher and U.~Wolff, {\it {How to calculate the elastic scattering matrix
  in two- dimensional quantum field theories by numerical simulation}},  {\em
  Nucl. Phys.} {\bf B339} (1990) 222--252.

\bibitem{Blossier:2009kd}
B.~Blossier, M.~Della~Morte, G.~von Hippel, T.~Mendes, and R.~Sommer, {\it {On
  the generalized eigenvalue method for energies and matrix elements in lattice
  field theory}},  {\em JHEP} {\bf 04} (2009) 094,
  [\href{http://arxiv.org/abs/0902.1265}{{\tt arXiv:0902.1265}}].

\bibitem{Dudek:2011b}
J.~J. Dudek, {\it {The lightest hybrid meson supermultiplet in QCD}},  {\em
  Phys. Rev.} {\bf D84} (2011) 074023,
  [\href{http://arxiv.org/abs/1106.5515}{{\tt arXiv:1106.5515}}].

\bibitem{Dudek:2012}
J.~J. Dudek and R.~G. Edwards, {\it {Hybrid Baryons in QCD}},  {\em Phys.Rev.}
  {\bf D85} (2012) 054016, [\href{http://arxiv.org/abs/1201.2349}{{\tt
  arXiv:1201.2349}}].

\bibitem{Edwards:2012fx}
{\bf Hadron Spectrum} Collaboration, R.~G. Edwards, N.~Mathur, D.~G. Richards,
  and S.~J. Wallace, {\it {Flavor structure of the excited baryon spectra from
  lattice QCD}},  {\em Phys. Rev.} {\bf D87} (2013), no.~5 054506,
  [\href{http://arxiv.org/abs/1212.5236}{{\tt arXiv:1212.5236}}].

\bibitem{Dudek:2013yja}
{\bf Hadron Spectrum} Collaboration, J.~J. Dudek, R.~G. Edwards, P.~Guo, and
  C.~E. Thomas, {\it {Toward the excited isoscalar meson spectrum from lattice
  QCD}},  {\em Phys. Rev.} {\bf D88} (2013), no.~9 094505,
  [\href{http://arxiv.org/abs/1309.2608}{{\tt arXiv:1309.2608}}].

\bibitem{Padmanath:2013zfa}
M.~Padmanath, R.~G. Edwards, N.~Mathur, and M.~Peardon, {\it {Spectroscopy of
  triply-charmed baryons from lattice QCD}},  {\em Phys. Rev.} {\bf D90}
  (2014), no.~7 074504, [\href{http://arxiv.org/abs/1307.7022}{{\tt
  arXiv:1307.7022}}].

\bibitem{Padmanath:2015jea}
M.~Padmanath, R.~G. Edwards, N.~Mathur, and M.~Peardon, {\it {Spectroscopy of
  doubly-charmed baryons from lattice QCD}},  {\em Phys. Rev.} {\bf D91}
  (2015), no.~9 094502, [\href{http://arxiv.org/abs/1502.01845}{{\tt
  arXiv:1502.01845}}].

\bibitem{Isgur:1991wq}
N.~Isgur and M.~B. Wise, {\it {Spectroscopy with heavy quark symmetry}},  {\em
  Phys. Rev. Lett.} {\bf 66} (1991) 1130--1133.

\bibitem{Moir:2013yfa}
{\bf Hadron Spectrum} Collaboration, G.~Moir, M.~Peardon, S.~M. Ryan, C.~E.
  Thomas, and L.~Liu, {\it {Excited spectroscopy of mesons containing charm
  quarks from lattice QCD}},  {\em PoS} {\bf LATTICE2013} (2014) 242,
  [\href{http://arxiv.org/abs/1312.1361}{{\tt arXiv:1312.1361}}].

\bibitem{Edwards:2004sx}
{\bf SciDAC} Collaboration, R.~G. Edwards and B.~Jo\'o, {\it {The Chroma
  software system for lattice QCD}},  {\em Nucl. Phys. B. Proc. Suppl.} {\bf
  140} (2005) 832, [\href{http://arxiv.org/abs/hep-lat/0409003}{{\tt
  hep-lat/0409003}}].

\bibitem{Clark:2009wm}
M.~A. Clark, R.~Babich, K.~Barros, R.~C. Brower, and C.~Rebbi, {\it {Solving
  Lattice QCD systems of equations using mixed precision solvers on GPUs}},
  {\em Comput. Phys. Commun.} {\bf 181} (2010) 1517--1528,
  [\href{http://arxiv.org/abs/0911.3191}{{\tt arXiv:0911.3191}}].

\bibitem{Babich:2010mu}
R.~Babich, M.~A. Clark, and B.~Jo\'o, {\it {Parallelizing the QUDA Library for
  Multi-GPU Calculations in Lattice Quantum Chromodynamics}},  in {\em
  International Conference for High Performance Computing, Networking, Storage
  and Analysis (SC)}, pp.~1--11, 2010.
\newblock \href{http://arxiv.org/abs/1011.0024}{{\tt arXiv:1011.0024}}.

\bibitem{ISC13Phi}
B.~Jo\'o, D.~Kalamkar, K.~Vaidyanathan, M.~Smelyanskiy, K.~Pamnany, V.~Lee,
  P.~Dubey, and W.~Watson, {\it {Lattice QCD on Intel Xeon Phi Coprocessors}},
  in {\em Supercomputing} (J.~Kunkel, T.~Ludwig, and H.~Meuer, eds.), vol.~7905
  of {\em Lecture Notes in Computer Science}, pp.~40--54.
\newblock Springer Berlin Heidelberg, 2013.

\bibitem{Osborn:2010mb}
J.~Osborn, R.~Babich, J.~Brannick, R.~Brower, M.~Clark, et~al., {\it {Multigrid
  solver for clover fermions}},  {\em PoS} {\bf LATTICE2010} (2010) 037,
  [\href{http://arxiv.org/abs/1011.2775}{{\tt arXiv:1011.2775}}].

\bibitem{Babich:2010qb}
R.~Babich, J.~Brannick, R.~Brower, M.~Clark, T.~Manteuffel, et~al., {\it
  {Adaptive multigrid algorithm for the lattice Wilson-Dirac operator}},  {\em
  Phys.Rev.Lett.} {\bf 105} (2010) 201602,
  [\href{http://arxiv.org/abs/1005.3043}{{\tt arXiv:1005.3043}}].

\end{thebibliography}\endgroup
\bibliographystyle{JHEP}

\end{document}